\documentclass[12pt]{article}
\usepackage{colortbl}
\usepackage{verbatim}
\usepackage{amssymb,amsfonts,amsmath,amsthm}
\usepackage{geometry}
\usepackage{alltt}
\usepackage{multirow}
\usepackage{graphicx}
\usepackage{subfigure}
\usepackage{setspace}
\usepackage{xcolor}
\usepackage{soul}
\usepackage{breqn}

\usepackage{booktabs} 
\usepackage{rotating}
\usepackage{bigstrut}

\usepackage{lscape}  
\usepackage{array} 
\usepackage[english]{babel}
\usepackage[sort, authoryear]{natbib} 
\usepackage{url}
\makeatletter
\renewcommand\@biblabel[1]{#1.} %from [1] to 1
\makeatother

\usepackage{bigstrut}
\usepackage[colorlinks,citecolor=blue,urlcolor=blue]{hyperref}
\usepackage{subfigure}
\usepackage{multirow}

\begin{document}
\title{Design-Based Inference for the AUC with Complex Survey Data}

\author{Amaia Iparragirre \footnote{Corresponding author: {{E-mail: amaia.iparragirre@ehu.eus}}, Address: Departamento de M\'etodos Cuantitativos. Universidad del Pa\'is Vasco UPV/EHU.} $^{1}$,
		Thomas Lumley$^2$,
        Irantzu Barrio$^{3,4}$\\
        \small{$^{1}$ Departamento de M\'etodos Cuantitativos.} 
        \small{Universidad del Pa\'is Vasco UPV/EHU}\\
        \small{$^{2}$ Department of Statistics. University of Auckland.}\\ 
         \small{$^{3}$ Departamento de Matem\'aticas.} 
        \small{Universidad del Pa\'is Vasco UPV/EHU}\\
        \small{$^{4}$ BCAM - Basque Center for Applied Mathematics}\\
}

\date{}
\maketitle

\begin{abstract}
 
Complex survey data are usually collected following complex sampling designs. Accounting for the sampling design is essential to obtain unbiased estimates and valid inference when analyzing complex survey data. The area under the receiver operating characteristic curve (AUC) is routinely used to assess the discriminative ability of predictive models for binary outcomes. However, valid inference for the AUC under complex sampling designs remains challenging. Although bootstrap techniques are widely applied under simple random sampling for variance estimation in this framework, traditional implementations do not account for complex designs.

In this work, we propose a design-based framework for AUC inference. In particular, replicate weights methods are used to construct confidence intervals and hypothesis test. The performance of replicate weights methods and the traditional non-design-based bootstrap for this purpose has been analized through an extensive simulation study. Design-based methods achieve coverage probabilities close to nominal levels and appropriate rejection rates under the null hypothesis. In contrast, the traditional non-design-based bootstrap method tends to underestimate the variance, leading to undercoverage and inflated rejection rates. Differences between methods decrease as the number of selected clusters per stratum increases.

An application to data from the National Health and Nutrition Examination Survey (NHANES) illustrates the practical relevance of the proposed framework. The methods have been incorporated into the \texttt{svyROC} R package.

\end{abstract}


\section{Introduction}

Complex sampling designs are widely implemented in epidemiological and population-based health studies. Notable examples include the National Health and Nutrition Examination Survey (NHANES), the Behavioral Risk Factor Surveillance System (BRFSS), the European Health Interview Survey (EHIS), and the Demographic and Health Surveys (DHS), all of which provide critical population-level health data across multiple countries. These surveys typically rely on stratification and clustering strategies, often implemented in one or multiple stages, to improve representativeness and data collection efficiency in large-scale population studies~\citep{Kaier1895, Kalton1983}. However, applying those sampling techniques for data collection also introduces complexities in the subsequent statistical analysis of the survey data, as traditional methods, which do not account for the survey design, can lead to biased estimates and invalid inferences~\citep{Skinner1989,Lumley2017a}.

In clinical and epidemiological studies, predictive models for binary outcomes are commonly fitted, and their discriminative performance is subsequently assessed~\citep{Akter2025,Wang2022,Zhang2017}. A standard summary measure of this discriminative performance is the area under the receiver operating characteristic (ROC) curve (AUC), which quantifies the model’s ability to distinguish between individuals with and without the event of interest. Inference for the AUC is well established under simple random sampling, for which bootstrap techniques are widely implemented \citep{Robin2011, Liu2006, Noma2021, Wu2016}. However, traditional implementations do not account for complex sampling schemes. 

As an alternative, in the context of complex survey data, replicate weights methods are widely used to generate partially independent samples mimmicking the complex sampling structure followed to collect the original sample~\citep{Heeringa2017, Wolter2007, Iparragirre2023a}. In a previous work, replicate weights methods were used to estimate the variance of an AUC estimator \citep{Yao2015}. In this work, we go a step further by proposing and evaluating the performance of replicate weight methods for constructing confidence intervals and conducting hypothesis tests. In particular, design-based approaches such as the Rescaling Bootstrap and Jackknife Repeated Replication have been considered, alongside the conventional (non-design-based) bootstrap. Using these methods, we construct confidence intervals for the AUC and perform hypothesis tests comparing two AUCs, considering both independent comparisons (the same model across different samples) and paired comparisons (different models within the same sample). The behaviour of these approaches is assessed through an extensive simulation study.

The rest of the paper is organized as follows. In Section~\ref{sec:methods}, we present the proposed methodology for AUC estimation and the construction of confidence intervals and hypothesis tests under complex sampling designs. Section~\ref{sec:simulation_study} describes an extensive simulation study evaluating the performance of the proposed approaches under various scenarios (additional scenarios and results are provided as Supplementary Material). Section~\ref{sec:application} presents an application of the proposed methodology to data from the NHANES survey. Finally, Section~\ref{sec:conclusions} concludes with a discussion and practical recommendations.

\section{Methods}\label{sec:methods}

This section describes the methodology proposed for the design-based inference of the AUC. Specifically, in Section~\ref{model_auc}, the logistic regression model and the estimation of its coefficients is described along with the AUC estimation of the model, in the context of complex survey data. Section~\ref{replicate_weights} describes the methods proposed in this work for variance estimation of the weighted AUC estimator. Finally, the proposal to define of confidence intervals and hypothesis tests is detailed in Section~\ref{ci_ht}.

\subsection{Logistic regression and AUC estimation with complex survey data}\label{model_auc}

Let $\pmb X=(1,X_1,\ldots,X_q)^T$ denote the vector of covariates, including an intercept, and $Y$ the dichotomous response variable (with $Y=1$ indicating the event of interest and $Y=0$ the non-event). The finite population of interest, denoted by $U$, consists of $N$ units with observed covariates $\pmb x_i$ and response values $y_i$, for all $i\in U$. The logistic regression model is defined through the \textit{logit} transformation as in eq.~\eqref{eq:logit},
\begin{equation}\label{eq:logit}
	logit\left[p(\pmb x_i)\right]=\ln\left[\dfrac{p(\pmb x_i)}{1-p(\pmb x_i)}\right]=\pmb x_i^T\pmb\beta,\quad\forall i\in U,
\end{equation}
where $\pmb\beta=(\beta_0,\beta_1,\ldots,\beta_q)^T$ denotes the vector of regression coefficients and $p(\pmb x_i)=P(Y=1|\pmb x_i)$.  The likelihood function in eq.~\eqref{eq:L} is maximized to compute the vector of regression coefficients $\pmb\beta$,
\begin{equation}\label{eq:L}
	L(\pmb\beta)=\prod_{i\in U} p(\pmb x_i)^{y_i}[1-p(\pmb x_i)]^{1-y_i}.
\end{equation}
Let us denote as ${\pmb\beta}^{\text{pop}}$ the vector of population coefficients obtained based on the maximization of the log-likelihood, and as $p_i^{\text{pop}}=p^{\text{pop}}(\pmb x_i)$ the corresponding probabilities of event, $\forall i\in U$. The discrimination ability of this population model can be determined as in eq. \eqref{eq:auc}:
\begin{equation}\label{eq:auc}
	AUC^{\text{pop}} = \dfrac{1}{{N_0N_1}}\sum_{i_0\in U_{Y=0}}\sum_{i_1\in U_{Y=1}} [I(p^{\text{pop}}_{i_0} <  p^{\text{pop}}_{i_1})+0.5I(p^{\text{pop}}_{i_0} = p^{\text{pop}}_{i_1})],
\end{equation}
where $U_{Y=0}$ and $U_{Y=1}$ are the subsets of the finite population $U$ formed by the units without ($Y=0$) and with ($Y=1$) the event of interest, respectively, $N_0$ and $N_1$ denote the number of units in $U_{Y=0}$ and $U_{Y=1}$, and $I(\cdot)$ is the indicator function.

However, in practice, the whole finite population will not be available for fitting the model, and regression coefficients and the corresponding AUC need to be estimated based on a sample $S$. Let us suppose that a sample $S$ is obtained following a complex sampling design, and the corresponding sampling weights ($w_i$) are available $\forall i \in S$, together with the covariates ($\pmb x_i$) and the outcome ($y_i$). The pseudo-likelihood function in eq.~\eqref{eq:PL} is commonly used for design-based estimation of $\pmb\beta$ under complex sampling~\citep{Binder1983}:
\begin{equation}\label{eq:PL}
	PL(\pmb\beta)=\prod_{i\in S} p(\pmb x_i)^{y_iw_i}[1-p(\pmb x_i)]^{(1-y_i)w_i}.
\end{equation}
Let us denote as $\hat{\pmb\beta}$ the vector of regression coefficients estimated based on eq.~\eqref{eq:PL}, and as $\hat p_i=\hat p(\pmb x_i)$ the corresponding estimated probabilities of event, $\forall i\in S$.

An AUC estimator that incorporates the sampling weights ($\widehat{AUC}_w$) as defined in eq.~\eqref{eq:aucw} has been proposed in the literature~\citep{Iparragirre2023} to estimate the discrimination ability of logistic regression models in the context of complex survey data based on $S$. Let $S_{Y=0}$ and $S_{Y=1}$ be the subsets of $S$ formed by units without ($Y=0$) and with ($Y=1$) the event of interest, respectively. Then,
\begin{equation}\label{eq:aucw}
	\widehat{AUC}_w=\dfrac{\sum_{i_0\in S_{Y=0}}\sum_{i_1\in S_{Y=1}} w_{i_0}w_{i_1}[I(\hat p_{i_0} < \hat p_{i_1})+0.5I(\hat p_{i_0} = \hat p_{i_1})]}{\sum_{i_0\in S_{Y=0}}\sum_{i_1\in S_{Y=1}} w_{i_0}w_{i_1}},
\end{equation}
provides a design-consistent point estimate of $AUC^{\text{pop}}$.

While $\widehat{AUC}_w$ provides a point estimate of the population AUC, valid statistical inference requires appropriate estimation of its variability under the complex sampling design.

\subsection{Variance estimation of the weighted AUC estimator}\label{replicate_weights}

The estimator of the AUC defined in eq.~\eqref{eq:aucw} is a nonlinear function of the estimated probabilities of event and the survey weights, and its variability is affected by the implemented complex sampling design. Consequently, deriving an analytical expression for its variance is not straightforward, motivating the use of replicate weights (design-based resampling methods) for variance estimation and inference \citep{Heeringa2017, Wolter2007}. In this work, we consider specific forms of Jackknife Repeated Replication (Section~\ref{sec:JKn}) and Bootstrap methods (Section~\ref{sec:boot}) to create partially independent subsets of the original sample $S$ while preserving the sampling design. These subsets are then used to estimate the variance of the AUC.

In order to set the notation, let $S^{(h,j)}\subset S$ denote the $j^{\text{th}}$ primary sampling unit (PSU, i.e., units sampled in the first stage of the sampling) from stratum $h$, $\forall j\in\{1,\ldots,a_h\}$ and $\forall h\in\{1,\ldots,H\}$, where $a_h$ indicates the total number of selected PSUs in stratum $h$. In particular, note that,
\begin{equation}
	S=\bigcup_{h=1}^H\bigcup_{j=1}^{a_h} S^{(h,j)}.
\end{equation}
In addition, let $S^{(h)}=\cup_{j=1}^{a_h} S^{(h,j)}$ denote the subset of the sample $S$ corresponding to stratum $h,\:\forall h\in\{1,\ldots,H\}$, being $H$ the total number of strata.

\subsubsection{Jackknife Repeated Replication}\label{sec:JKn}

In the Jackknife Repeated Replication method \citep{Heeringa2017, Wolter2007} (JKn, hereinafter), new sets are created by systematically leaving out one PSU at a time, so that each PSU, $S^{(h,j)}$, is excluded once, $\forall h\in\{1,\ldots,H\},\:\forall j\in\{1,\ldots,a_h\}$. The sampling weights are then adjusted so that the remaining units in each set properly represent the entire finite population $U$. In particular, let us suppose that PSU $S^{(h,j)}$ is removed from the original set to form the new set. The new set will be denoted as $S^{-(h,j)}=S-S^{(h,j)},\:\forall h\in\{1,\ldots,H\},\:\forall j\in\{1,\ldots,a_h\}$. Then, the corresponding replicate weights are defined as in eq. \eqref{eq:rw_JKn}:
\begin{equation}\label{eq:rw_JKn}
	w_i^{-(h,j)}=\left\{
	\begin{array}{cl}
		0, & \text{if } i\in S^{(h,j)},\\
		w_i\cdot\dfrac{a_h}{a_h-1}, & \text{if } i\in S^{(h)}\text{ but } i\notin S^{(h,j)},\\
		w_i, & \text{if } i\notin S^{(h)},
	\end{array}
	\right.\quad \forall i\in S.
\end{equation}
In this way, a total of $a=\sum_{h=1}^H a_h$ new sets are defined.

The AUC is computed in each set $S^{-(h,j)}$ by following eq.~\eqref{eq:aucw} and using the replicate weights defined in eq.~\eqref{eq:rw_JKn}, as shown in eq. \eqref{eq:aucw_jkn}:
\begin{equation}\label{eq:aucw_jkn}
	\widehat{AUC}_w^{-(h,j)}=\dfrac{\sum_{i_0\in S^{-(h,j)}_{Y=0}}\sum_{i_1\in S^{-(h,j)}_{Y=1}} w^{-(h,j)}_{i_0}w^{-(h,j)}_{i_1}[I(\hat p_{i_0} < \hat p_{i_1})+0.5I(\hat p_{i_0} = \hat p_{i_1})]}{\sum_{i_0\in S^{-(h,j)}_{Y=0}}\sum_{i_1\in S^{-(h,j)}_{Y=1}} w^{-(h,j)}_{i_0}w^{-(h,j)}_{i_1}}.
\end{equation}
Finally, the variance of $\widehat{AUC}_w$ is estimated using eq.~\eqref{eq:var_JKn} \citep{Wolter2007, Yao2015}:
\begin{equation}\label{eq:var_JKn}
	\widehat{var}_{JKn}(\widehat{AUC}_w)=\sum_{h=1}^H \dfrac{a_h-1}{a_h}\sum_{j=1}^{a_h}(\widehat{AUC}_w^{-(h,j)}-\widehat{AUC}_w)^2.
\end{equation} 

\subsubsection{Bootstrap}\label{sec:boot}

Based on bootstrap methods, variance is estimated by generating $B$ resamples of the original sample ($S^{(b)},\:\forall b\in\{1,\ldots,B\}$) drawn with replacement. In this work, we consider three variants of the bootstrap: two design-based versions of the Rescaling Bootstrap that account for the survey design when generating the resamples, and the traditional (non-design-based) bootstrap. These approaches are described in detail below.

In the Rescaling Bootstrap \citep{Rao1988} (RB, hereinafter), the resampling is performed by randomly selecting $a_h-1$ PSUs with replacement from each stratum $h$, $\forall h\in\{1,\ldots,H\}$, to generate each bootstrap sample $S^{RB(b)},\:b\in\{1,\ldots,B\}$. The sampling weights are then recalculated as indicated in eq.~\eqref{eq:rw_RB}:
\begin{equation}\label{eq:rw_RB}
	w_i^{RB(b)}=w_i\cdot\dfrac{a_h}{a_h-1}\cdot k_{(h,j)}^{(b)},\quad \forall i\in S^{(h,j)},\:\forall h\in\{1,\ldots,H\},\:\forall j\in\{1,\ldots,a_h\}.
\end{equation}
where $k_{(h,j)}^{(b)}$ denotes the number of times PSU $S^{(h,j)}$ is selected to form the resample $S^{RB(b)}$(note that $k_{(h,j)}^{(b)}\geq 0$, and $k_{(h,j)}^{(b)}=0$ if the PSU is not selected in that resample).

Similarly, another bootstrap variant (RBn, hereinafter) selects $a_h$ PSUs from each stratum~\citep{Canty1999} $h\in\{1,\ldots,H\}$, rather than $a_h-1$ as in RB. In this case, the replicate weights corresponding to the resample $S^{RBn(b)},\:\forall b\in\{1,\ldots,B\}$ are recalculated as in eq. \eqref{eq:rw_RBn}:
\begin{equation}\label{eq:rw_RBn}
	w_i^{RBn(b)}=w_i\cdot k_{(h,j)}^{(b)},\quad \forall i\in S^{(h,j)},\:\forall h\in\{1,\ldots,H\},\:\forall j\in\{1,\ldots,a_h\}.
\end{equation}

In addition to the design-based methods, we consider the traditional bootstrap \citep{Efron1979,Robin2011} (trB), which does not account for the complex sampling design during resampling. In this approach, $n$ units are randomly sampled with replacement from $S$ to form each bootstrap sample $S^{trB(b)}$, ignoring the stratum and/or the cluster to which each unit belongs. To maintain consistency with the notation used in RB and RBn, the sampling weights considered in $S^{trB(b)}$ are expressed as in eq.~\eqref{eq:rw_trB},
\begin{equation}\label{eq:rw_trB}
	w_i^{trB(b)}=w_i\cdot k_{i}^{(b)}, \quad\forall i\in S.
\end{equation}
where $k_{i}^{(b)}$ indicates the number of times that unit $i\in S$ is selected in bootstrap resample $S^{trB(b)},\:\forall b\in\{1,\ldots,B\}$.

In any of the bootstrap methods defined above (denoted generically as $boot\in\{RB,RBn,trB\}$ for ease of notation), the AUC of the model is computed in each resample $S^{boot(b)}$, as defined in eq.~\eqref{eq:aucw_boot}, $\forall{boot\in\{RB,RBn,trB\}}$ and $\forall b\in\{1,\ldots,B\}$: 
\begin{equation}\label{eq:aucw_boot}
	\widehat{AUC}_w^{boot(b)}=\dfrac{\sum_{i_0\in S^{boot(b)}_{Y=0}}\sum_{i_1\in S^{boot(b)}_{Y=1}} w^{boot(b)}_{i_0}w^{boot(b)}_{i_1}[I(\hat p_{i_0} < \hat p_{i_1})+0.5I(\hat p_{i_0} = \hat p_{i_1})]}{\sum_{i_0\in S^{boot(b)}_{Y=0}}\sum_{i_1\in S^{boot(b)}_{Y=1}} w^{boot(b)}_{i_0}w^{boot(b)}_{i_1}},
\end{equation}
where $S^{boot(b)}_{Y=0}$ and $S^{boot(b)}_{Y=1}$ are the subset of $S^{boot(b)}$ formed by the units without and with the event of interest, respectively.\\

The variance of $\widehat{AUC}_w$ is then estimated as in eq.~\eqref{eq:var_boots}~\citep{Wolter2007}, $\forall boot\in\{RB, RBn, trB\}$:
\begin{equation}\label{eq:var_boots}
	\widehat{var}_{boot}(\widehat{AUC}_w)=\dfrac{1}{B-1}\sum_{b=1}^B\left(\widehat{AUC}^{boot(b)}_{w}-\overline{\widehat{AUC}^{boot(b)}_{w}}\right)^2,
\end{equation}
where,
\begin{equation}
	\overline{\widehat{AUC}^{boot(b)}_{w}}=\dfrac{1}{B}\sum_{b=1}^B \widehat{AUC}^{boot(b)}_{w}.
\end{equation}

\subsection{Confidence intervals and hypothesis testing}\label{ci_ht}

In this section, the construction of confidence intervals and hypothesis tests is defined. In particular, the definition of confidence intervals (CIs) is provided in Section \ref{sec:CI} and the hypothesis tests are detailed in Section \ref{sec:HT}. We denote the confidence level by $1-\alpha$ (i.e., $100\cdot(1-\alpha)\%$ CIs) and the significance level for hypothesis tests by $\alpha$.

\subsubsection{Confidence intervals for the AUC}\label{sec:CI}

This section describes the construction of confidence intervals for the AUC using the JKn and bootstrap (RB, RBn, and trB) methods.

\begin{itemize}
	\item \textbf{JKn:} For the JKn method, the confidence intervals are defined as shown in eq. \eqref{eq:CI_JKn}:
	\begin{equation}\label{eq:CI_JKn}
		\mathcal{I}_{AUC,\:JKn}^{1-\alpha}=\left(\widehat{AUC}_w\pm z_{\alpha/2}\sqrt{\widehat{var}_{JKn}(\widehat{AUC}_w)}\right),
	\end{equation}
	where $\widehat{var}_{JKn}(\widehat{AUC}_w)$ is given in eq. \eqref{eq:var_JKn}, $\widehat{AUC}_w$ is the point estimate defined in eq. \eqref{eq:aucw}, and $z_{\alpha/2}$ denotes the upper $\alpha/2$ critical value of the standard normal distribution, i.e., the value such that $P(Z \geq z_{\alpha/2})=\alpha/2$ for $Z\sim N(0,1)$.
	
	\item \textbf{Bootstrap (RB, RBn, trB):} Similarly, for any of the bootstrap methods ($\forall boot\in\{RB,RBn,trB\}$), the confidence intervals can be defined as shown in eq.~\eqref{eq:CI_boot}:
	\begin{equation}\label{eq:CI_boot}
		\mathcal{I}_{AUC,\:boot}^{1-\alpha}=\left(\widehat{AUC}_w\pm z_{\alpha/2}\sqrt{\widehat{var}_{boot}(\widehat{AUC}_w)}\right),
	\end{equation}
	where $\widehat{var}_{boot}(\widehat{AUC}_w)$ is given in eq. \eqref{eq:var_boots}, $\widehat{AUC}_w$ is the point estimate defined in eq. \eqref{eq:aucw}, and $z_{\alpha/2}$ denotes the upper $\alpha/2$ critical value of the standard normal distribution as indicated above. Another widely implemented way to define confidence intervals for any of the bootstrap methods ($\forall boot\in\{RB,RBn,trB\}$), is using the quantiles of the empirical distribution of the bootstrap AUC estimates \citep{Efron1981,Robin2011} as shown in eq.~\eqref{eq:CI_boot_q},
	\begin{equation}\label{eq:CI_boot_q}
		\mathcal{I}_{AUC,\:boot}^{1-\alpha}=\left(\theta_{\alpha/2},\:\theta_{1-\alpha/2}\right),
	\end{equation}
	where $\theta_{\alpha/2}$ and $\theta_{1-\alpha/2}$ denote the $\alpha/2$ and $1-\alpha/2$ quantiles of the empirical distribution of the $B$ bootstrap AUC estimates defined in eq.~\eqref{eq:aucw_boot}. 
\end{itemize}

\subsubsection{Hypothesis Testing for the comparison of two AUCs}\label{sec:HT}

Let $AUC^{\text{pop}}_{(1)}$ and $AUC^{\text{pop}}_{(2)}$ denote two population AUCs, and define their difference as in eq.~\eqref{eq:Dpop}:
\begin{equation}\label{eq:Dpop}
	D^{\text{pop}} = AUC^{\text{pop}}_{(1)} - AUC^{\text{pop}}_{(2)}.
\end{equation}
The hypothesis test of interest is given in eq.~\eqref{eq:HT}:
\begin{equation}\label{eq:HT}
	\left\{
	\begin{array}{cc}
		H_0: & D^{\text{pop}} = 0,\\
		H_1: & D^{\text{pop}} \neq 0.
	\end{array}
	\right.
\end{equation}

Let $\widehat{AUC}_{w(1)}$ and $\widehat{AUC}_{w(2)}$ be the corresponding point estimates of the population AUCs computed as in eq.~\eqref{eq:aucw}. The estimator of $D^{\text{pop}}$ is defined in eq.~\eqref{eq:Dhat},
\begin{equation}\label{eq:Dhat}
	\widehat D = \widehat{AUC}_{w(1)} - \widehat{AUC}_{w(2)}.
\end{equation}

To test the null hypothesis, we consider the Wald-type statistic defined in eq.~\eqref{eq:z_statistic}:
\begin{equation}\label{eq:z_statistic}
	Z=\dfrac{\widehat{D}-D^{\text{pop}}}{\sqrt{\widehat{var}(\widehat{D})}},
\end{equation}
which, under the null hypothesis, is assumed to be asymptotically standard normal $N(0,1)$ (note also that under $H_0$, $D^{\text{pop}}=0$).

However, the variance estimation of $\widehat{D}$ depends on whether the two AUCs are independent or paired. In addition, as detailed above, the variance can be estimated by means of any of the methods previously described ($\forall m\in\{JKn, RB, RBn, trB\}$). This process, for all the methods $m$ considered, is described below, for independent AUCs in eq.~\eqref{eq:z_unpaired} and for paired AUCs in eq.~\eqref{eq:z_paired_final}. 

Let $z_m$ denote the value of the test statistic obtained based on method $m$, $\forall m\in\{JKn, RB, RBn, trB\}$, for the comparison of either independent or paired AUCs. Once $z_m$ is computed, the corresponding two-sided p-value is obtained as in eq.~\eqref{eq:pvalue}:
\begin{equation}\label{eq:pvalue}
	p = 2P(Z > |z_m|),\quad Z\sim N(0,1).
\end{equation}
The null hypothesis is rejected at significance level $\alpha$ if $p < \alpha$.

Below, we detail the process for computing the test statistic for the comparison of both independent and paired AUCs.
\\~\\

\noindent\textbf{(a) Statistic value ($z_m$) for two independent AUCs}\\

\noindent Let $S_{(1)}$ and $S_{(2)}$ be two independent samples. A model is fitted separately to each sample following eq.~\eqref{eq:PL}, yielding the estimated coefficient vectors $\hat{\pmb\beta}_{(1)}$ and $\hat{\pmb\beta}_{(2)}$, and the corresponding AUC estimates $\widehat{AUC}_{w(1)}$ and $\widehat{AUC}_{w(2)}$, computed as in eq.~\eqref{eq:aucw}. Since the two samples are independent, the covariance between the two AUC estimators is null. Consequently, for any method $m \in \{JKn, RB, RBn, trB\}$, the test statistic is computed as in eq.~\eqref{eq:z_unpaired}:
\begin{equation}\label{eq:z_unpaired}
	z_m
	= \dfrac{\widehat{D}}{\sqrt{\widehat{var}_m(\widehat{D})}}
	= \dfrac{\widehat{D}}
	{\sqrt{\widehat{var}_m(\widehat{AUC}_{w(1)}) + \widehat{var}_m(\widehat{AUC}_{w(2)})}},
\end{equation}
where, the $\widehat{var}_m(\widehat{AUC}_{w(1)})$ and $\widehat{var}_m(\widehat{AUC}_{w(2)})$ are computed using the methods described in eq.~\eqref{eq:var_JKn} for $m=JKn$ and in eq.~\eqref{eq:var_boots} for the bootstrap approaches, $\forall m\in\{RB, RBn, trB\}$. \\~\\

\noindent\textbf{(b) Statistic value ($z_m$) for two paired AUCs}\\

\noindent Let $S$ be a sample with observations on the response variable $Y$ and covariates $\pmb X$. Let $\pmb X^{(1)}$ and $\pmb X^{(2)}$ be two subsets of $\pmb X$ used to fit two different models, for which coefficients $\hat{\pmb\beta}_{(1)}$ and $\hat{\pmb\beta}_{(2)}$ are estimated, yielding estimated probabilities $\hat p_i^{(1)}$ and $\hat p_i^{(2)}$, $\forall i \in S$, and corresponding weighted AUCs, $\widehat{AUC}_{w(1)}$ and $\widehat{AUC}_{w(2)}$, following eq.~\eqref{eq:aucw}. 

Since both AUCs are estimated from the same sample, they are correlated. Therefore, the variance of $\widehat D$ is estimated by computing AUC differences across replicates and assessing their variability, as detailed below.

\begin{itemize}
	\item \textbf{JKn:} For each replicate set $S^{-(h,j)}$ described in Section \ref{sec:JKn}, AUCs $\widehat{AUC}_{w(1)}^{-(h,j)}$ and $\widehat{AUC}_{w(2)}^{-(h,j)}$ are computed from the corresponding probabilities. The difference for each replicate is given in eq.~\eqref{eq:D_JKn_paired}:
	\begin{equation}\label{eq:D_JKn_paired}
		\widehat D^{-(h,j)} = \widehat{AUC}_{w(1)}^{-(h,j)} - \widehat{AUC}_{w(2)}^{-(h,j)}, \quad \forall h \in \{1,\ldots,H\}, \forall j \in \{1,\ldots,a_h\}.
	\end{equation}
	The variance of $\widehat D$ is then estimated as in eq.~\eqref{eq:var_JKn_D_paired}:
	\begin{equation}\label{eq:var_JKn_D_paired}
		\widehat{var}_{JKn}(\widehat D)=\sum_{h=1}^H \frac{a_h-1}{a_h} \sum_{j=1}^{a_h} (\widehat D^{-(h,j)} - \widehat D)^2.
	\end{equation}
	
	\item \textbf{Bootstrap (RB, RBn, trB):} For each resample $S^{boot(b)}$  described in Section \ref{sec:boot}, AUCs $\widehat{AUC}_{w(1)}^{boot(b)}$ and $\widehat{AUC}_{w(2)}^{boot(b)}$ are computed $\forall boot\in\{RB, RBn, trB\}$, and their difference is calculated as in eq.~\eqref{eq:D_boot_paired}:
	\begin{equation}\label{eq:D_boot_paired}
		\widehat D^{boot(b)} = \widehat{AUC}_{w(1)}^{boot(b)} - \widehat{AUC}_{w(2)}^{boot(b)}, \quad \forall b \in \{1,\ldots,B\}.
	\end{equation}
	The variance of $\widehat D$ is estimated as in eq.~\eqref{eq:var_boot_D_paired}:
	\begin{equation}\label{eq:var_boot_D_paired}
		\widehat{var}_{boot}(\widehat D) = \frac{1}{B-1} \sum_{b=1}^B (\widehat D^{boot(b)} - \overline{\widehat D^{boot(b)}})^2, \quad
		\text{where}\quad\overline{\widehat D^{boot(b)}} = \frac{1}{B} \sum_{b=1}^B \widehat D^{boot(b)}.
	\end{equation}
\end{itemize}

Finally, the test statistic for each method is calculated as in eq.~\eqref{eq:z_paired_final}:
\begin{equation}\label{eq:z_paired_final}
	z_m = \frac{\widehat D}{\sqrt{\widehat{var}_m(\widehat D)}}, \quad \forall m \in \{JKn, RB, RBn, trB\},
\end{equation}
where $\widehat{var}_m(\widehat D)$ is calculated as in eq. \eqref{eq:var_JKn_D_paired} for $m=JKn$ and as in eq.~\eqref{eq:var_boot_D_paired}, $\forall m\in\{RB,RBn,trB\}$.

\section{Simulation study}\label{sec:simulation_study}

An extensive simulation study has been conducted to evaluate the performance of the JKn and the different bootstrap methods for the construction of confidence intervals and hypothesis testing for the AUC in the context of complex survey data. Section \ref{sec:data_generation} describes the data generation process and the considered scenarios. Section \ref{sec:results} summarizes the main results. Finally, in Section \ref{sec:additional_scenarios} further scenarios for a broader comparison of paired AUCs are analyzed and discussed.

\subsection{Data generation and scenarios}\label{sec:data_generation}

This section describes the data generation process followed in the simulation study and the considered scenarios. Depending on whether the objective is to define a confidence interval or to conduct a hypothesis test for the comparison of two independent or paired AUCs, different numbers of populations are generated and different types of models are fitted. In all cases, the generated populations are subsequently sampled according to a complex sampling design. Below, we describe both the population generation process under the different scenarios in Section~\ref{sec:population_generation} and the sampling design used to obtain the samples in Section~\ref{sec:sampling_design}.

\subsubsection{Finite population generation}\label{sec:population_generation}

The finite populations were generated as follows. A probability of event or prevalence of $P(Y=1)=0.5$ was predefined. Each population consists of $N=100\,000$ units, with information on covariates ($\pmb X$, 4 variables) and design variables ($\pmb Z$, 6 variables), generated conditionally on $Y$ from a multivariate normal distribution as in eq.~\eqref{eq:covariate_generation} (for ease of notation, let $\pmb X^*$ denote the combined vector of covariates and design variables, i.e., $\pmb X^*=(\pmb X,\pmb Z)$):
\begin{equation}\label{eq:covariate_generation}
	\pmb X^* \mid Y = 0 \sim N(\pmb \mu_{Y=0}, \Sigma_{Y=0}), \quad\text{and}\quad 	\pmb X^* \mid Y = 1 \sim N(\pmb \mu_{Y=1}, \Sigma_{Y=1}).
\end{equation}
The variance-covariance matrices were defined as in eq. \eqref{eq:var_covar},
\begin{equation}\label{eq:var_covar}
	\Sigma_{Y=0} = \Sigma_{Y=1} = (1-\gamma) \cdot I_{10 \times 10} + \gamma \cdot J_{10 \times 10},
\end{equation}
where $\gamma = 0.15$, $I_{10 \times 10}$ is the $10 \times 10$ identity matrix, and $J_{10 \times 10}$ is a matrix of ones of the same dimension. The mean vector for the non-event group $\pmb \mu_{Y=0}$ was set to zero, whereas $\pmb \mu_{Y=1}$ varies across scenarios as detailed in Table~\ref{tab:data_generation_summary}.

Under the assumptions indicated above, the logistic regression model is correctly specified and the vector of regression coefficients $\pmb \beta^* = (\pmb \beta_X, \pmb \beta_Z)$ is known \citep{Iparragirre2019}. The finite population was then generated according to the following steps:
\begin{enumerate}
	\item[P1] For $i=1,\ldots,N/2$ generate $\pmb x^*_i \sim N(\pmb \mu_{Y=0}, \Sigma_{Y=0})$ and for $i=N/2+1,\ldots,N$ $\pmb x^*_i \sim N(\pmb \mu_{Y=1}, \Sigma_{Y=1})$.
	\item[P2] Compute the probability of event $p(\pmb x^*_i)$ for each unit using $\pmb\beta^*$.
	\item[P3] Generate the binary response $y_i \sim \text{Bernoulli}[p(\pmb x^*_i)]$, independently for each unit.
	\item[P4] Define the survey design in the population:  
	\begin{enumerate}
		\item The coefficients associated with the design variables, $\pmb\beta_Z$, were derived from the full coefficient vector $\pmb\beta^*$.  
		\item Sort the data by $\pmb z_i \pmb \beta_Z$ for all $i=1,\dots,N$.  
		\item Define $H=5$ strata by partitioning the population into equal-sized sets.  
		\item Within each stratum $h$, define $A_h = 20$ clusters of equal size, yielding a total of $A=100$ clusters with $N_{h,j^*} = 1\,000$ units per cluster, for $h=1,\dots,H$ and $j^*=1,\dots,A_h$.
	\end{enumerate}
\end{enumerate}

After generating the finite population $U$, the model was fitted following eq. \eqref{eq:L} to the whole population using only the covariates $\pmb X$ (excluding the design variables $\pmb Z$). The resulting AUC was computed following eq. \eqref{eq:auc} and taken as the finite population AUC, denoted by $AUC^{\text{pop}}$ in Table~\ref{tab:data_generation_summary}.

For each simulation scenario, a different number of finite populations and fitted models were considered depending on the inferential objective. Specifically, one population and one model were used for confidence interval estimation; two populations and one model for hypothesis testing of independent AUCs; and one population with two different models for paired AUC comparisons. These settings, together with the corresponding population AUC values, are summarized in Table \ref{tab:data_generation_summary}.

\begin{landscape}
\begin{center}
	\begin{table}[h!]%
		\caption{Summary of the scenarios considered in the simulation study. For each scenario, the type of analysis (confidence interval, CI; hypothesis test, HT), the number of finite populations generated, the mean vector $\pmb\mu_{Y=1}^T$ of the covariates for units with $Y=1$, the covariates considered in the model, and the corresponding population AUC are reported.}\label{tab:data_generation_summary}
		\begin{tabular}{cccccc}
			\toprule
			Scenario & Contrast & Population & $\pmb\mu_{Y=1}^T$ & Covariates & $AUC^{\text{pop}}$\\
			\midrule
			1 & CI & $(1)$  &  $(0.7,0.7,0.7,0.7,0.7,0.7,0.7,0.7,0.7,0.7)$  &  $X_1,X_2,X_3,X_4$  & 0.7951\\
			\midrule
			2 & HT & $(1)$  &  $(0.7,0.7,0.7,0.7,0.7,0.7,0.7,0.7,0.7,0.7)$  &  $X_1,X_2,X_3,X_4$  & 0.7951\\
			&independent$^{\rm *}$ & $(2)$  &  $(0.7,0.7,0.7,0.7,0.7,0.7,0.7,0.7,0.7,0.7)$  &  $X_1,X_2,X_3,X_4$  & 0.7941\\
			\midrule
			3 & HT & $(1)$  &  $(0.7,0.7,0.7,0.7,0.7,0.7,0.7,0.7,0.7,0.7)$  &  $X_1,X_2,X_3,X_4$  & 0.7951\\
			&independent & $(2)$  &  $(0.7,0.7,0.7,1.2,0.7,0.7,0.7,0.7,0.7,0.7)$  &  $X_1,X_2,X_3,X_4$  & 0.8474\\
			\midrule
			4 & HT & $(1)$  &  $(0.7,0.7,0.7,0.7,0.7,0.7,0.7,0.7,0.7,0.7)$  &  $X_1,X_2,X_3$  & 0.7755\\
			&paired &   &    &  $X_1,X_2,X_4$  & 0.7743\\
			\midrule
			5 & HT & $(1)$  &  $(0.7,0.7,0.9,1.1,0.7,0.7,0.7,0.7,0.7,0.7)$  &  $X_1,X_2,X_3$  & 0.7991\\
			&paired &   &   &  $X_1,X_2,X_4$  & 0.8237 \\
			\bottomrule
		\end{tabular}
		\vspace{-0.5cm}\item[$^{\rm *}$] Different seeds have been used to generate two different finite populations.
	\end{table}
\end{center}
\end{landscape}

\subsubsection{Sampling design and weights}\label{sec:sampling_design}

The sampling of the generated finite populations was carried out following a two-stage stratified cluster sampling design: in the first stage, $a_h$ clusters were selected from each stratum $h\in\{1,\ldots,H\}$, and in the second stage, $n_{h,j}$ units were sampled within each selected cluster $j\in\{1,\ldots,a_h\}$. 

Two different sample sizes were considered, and four different numbers of clusters per stratum were evaluated. This allowed us to study the effect of increasing the sample size as well as the effect of increasing the number of clusters while keeping the total sample size approximately constant. Table~\ref{tab:sampling} summarizes the number of clusters selected per stratum ($a_h\in\{2,4,8,10\}$) and the corresponding number of sampled units in each scenario. In particular, the smallest sample sizes ($n_1$) resulted in $1\,680$–$1\,740$ units, while the second setting doubled these sizes ($n_2$). 

Sampling weights for each unit were calculated according to the standard formula for two-stage stratified cluster designs, as shown in eq.~\eqref{eq:sampling_weight}:
\begin{equation}\label{eq:sampling_weight}
	w_i=\dfrac{A_h}{a_h}\cdot\dfrac{N_{h,j}}{n_{h,j}},\quad \forall i\in S^{(h,j)}, \forall h\in\{1,\ldots,H\},\forall j\in\{1,\ldots,a_h\}.
\end{equation}

\begin{figure}[h!]
	\centering
	\includegraphics[width=10cm]{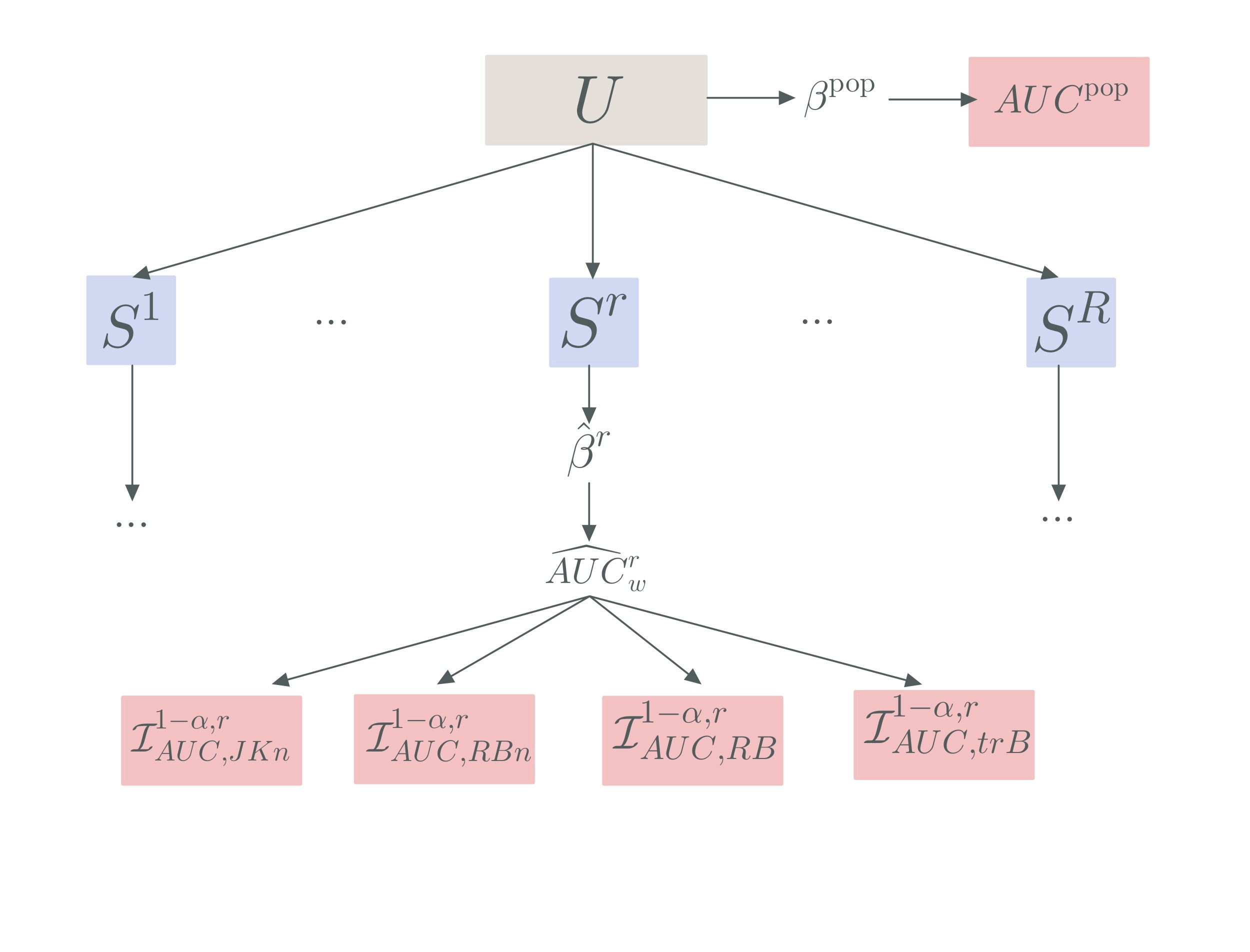}
	\caption{Graphical summary of the simulation set-up for confidence intervals.}
	\label{fig:sim_ci}
\end{figure}

\begin{figure}[h!]
	\includegraphics[width=0.5\textwidth]{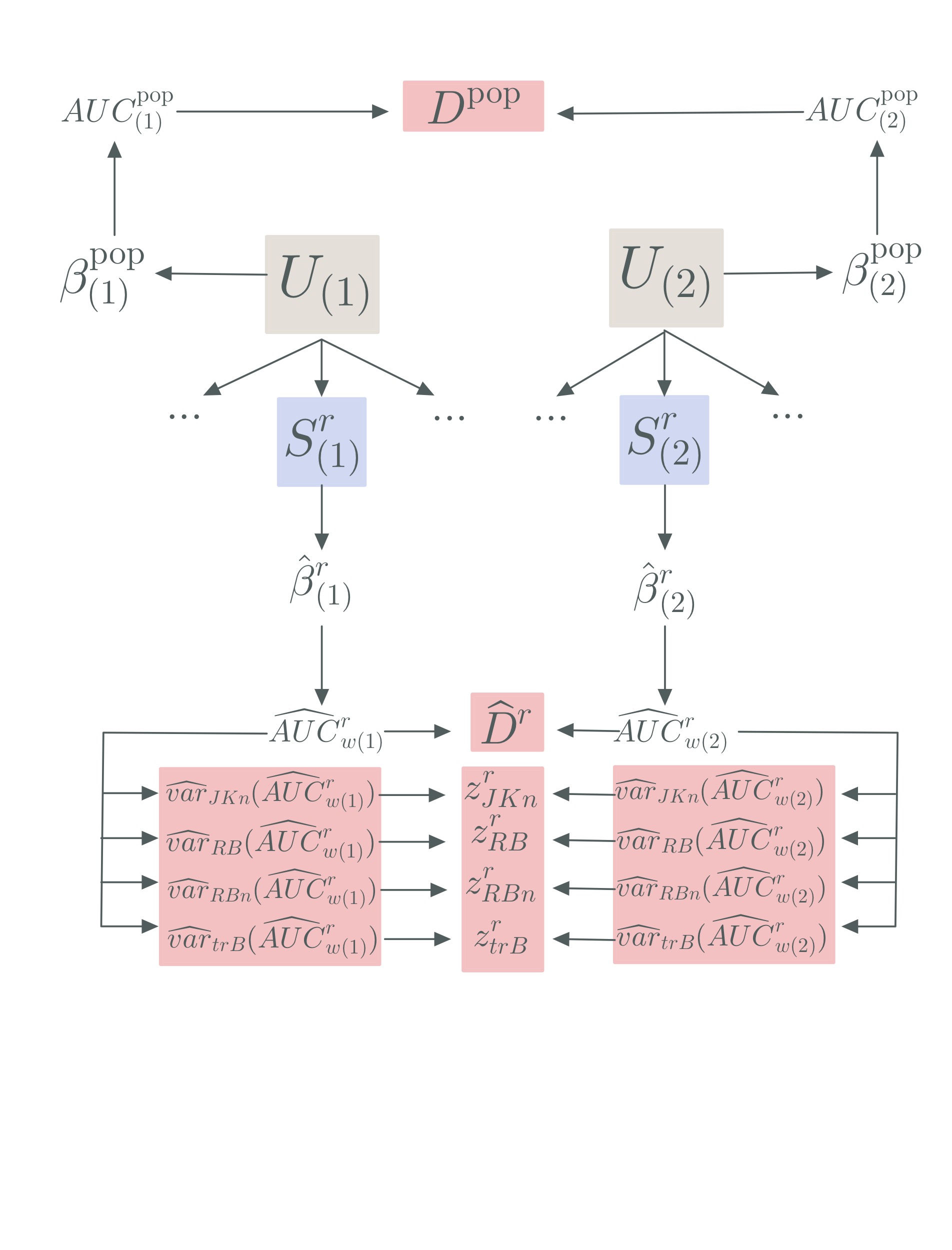}
	\includegraphics[width=0.5\textwidth]{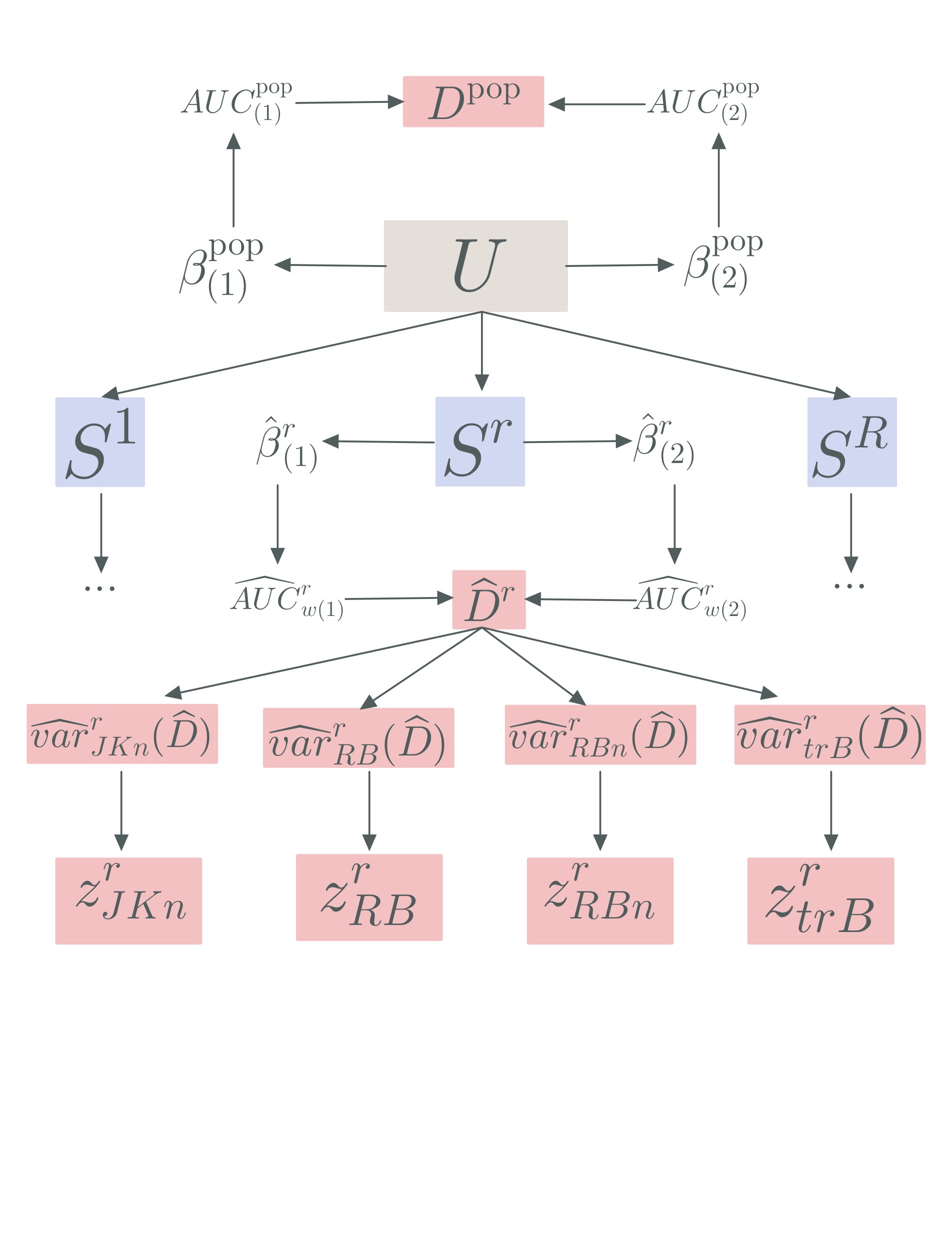}
	\caption{Graphical summary of the simulation set-up for hypothesis tests. The left figure illustrates the process followed for the comparison of two independent AUCs, while the right figure depicts the paired AUC comparison.}
	\label{fig:sim_ht}
\end{figure}

In each scenario, each finite population was sampled $R$ times according to the complex sampling design described above. For every sample $S^r,\:\forall r=1,\ldots,R$, the corresponding model or models  (depending on the scenario, as indicated in Table~\ref{tab:data_generation_summary}) were fitted to the sampled data by maximizing the pseudo-likelihood function in eq.~\eqref{eq:PL}, the estimated coefficients $\hat{\pmb\beta}^r$ were obtained and the corresponding $\widehat{AUC}_w^r$ values were obtained following eq.~\eqref{eq:aucw}. Subsequently, the variance estimators described in Section~\ref{replicate_weights} were applied to construct confidence intervals for the AUC or to compute the test statistic value for hypothesis testing as indicated in Section~\ref{ci_ht}, depending on the analysis considered. A graphical summary of this simulation process is provided in Figure~\ref{fig:sim_ci} (confidence intervals) and Figure~\ref{fig:sim_ht} (hypothesis tests). It should be noted that, in Scenario 2, two populations were generated (as indicated in Table~\ref{tab:data_generation_summary}) to ensure consistency across all scenarios involving the comparison of independent AUCs. However, in this particular case, an equivalent setup could be obtained by generating a single population and drawing two independent samples from it.

The simulation study was based on $R=500$ runs, with $B=1\,000$ bootstrap replicates implemented within each run and significance levels $\alpha \in \{0.01, 0.05, 0.1\}$ considered for both confidence interval construction and hypothesis testing. Confidence interval performance was evaluated through estimated coverage probabilities, defined as the proportion of simulation runs in which the population AUC ($AUC^{\text{pop}}$) was contained within the estimated interval as shown in eq.~\eqref{eq:coverage}:
\begin{equation}\label{eq:coverage}
	c_m^{1-\alpha}=\dfrac{1}{R}\sum_{r=1}^R I(AUC^{\text{pop}}\in\mathcal{I}^{1-\alpha,r}_{AUC,m}),\quad\forall m\in\{JKn,RB,RBn,trB\}.
\end{equation}
For hypothesis testing, performance was assessed through rejection rates, computed as the proportion of runs in which the null hypothesis was rejected at level $\alpha$ as defined in eq.~\eqref{eq:rejection_rate}:
\begin{equation}\label{eq:rejection_rate}
	rr_m^{\alpha}=\dfrac{1}{R}\sum_{r=1}^R I(p_m^r<\alpha),\quad\forall m\in\{JKn,RB,RBn,trB\},
\end{equation}
where $p_m^r$ indicates the p-value corresponding to the statistic value $z^r_m$ $\forall r\in\{1,\ldots,R\}$, calculated as in eq.~\eqref{eq:pvalue}.


\begin{landscape}
	\begin{table}[h]%
		\small
		\centering
		\caption{Summary of the sampling schemes used in the simulation study. 
			For each number of clusters per stratum $a_h\in\{2,4,8,10\}$, the number of units sampled within each cluster ($\forall j\in\{1,\ldots,a_h\}$) is reported for the two different overall sample sizes considered.}\label{tab:sampling}
		\begin{tabular}{ccc}
			\toprule
			$a_h$ & Sample size ${(n_1)}$ & Sample size ${(n_2)}$\\
			\midrule
			2 & $n_{1,j}^{(1)}=300;\:n_{2,j}^{(1)}=100;\:n_{3,j}^{(1)}=50;\:n_{4,j}^{(1)}=100;\:n_{5,j}^{(1)}=300$ & $n_{1,j}^{(2)}=600;\:n_{2,j}^{(2)}=200;\:n_{3,j}^{(2)}=100;\:n_{4,j}^{(2)}=200;\:n_{5,j}^{(2)}=600$ \\
			4 & $n_{1,j}^{(1)}=150;\:n_{2,j}^{(1)}=50;\:n_{3,j}^{(1)}=25;\:n_{4,j}^{(1)}=50;\:n_{5,j}^{(1)}=150$ & $n_{1,j}^{(2)}=300;\:n_{2,j}^{(2)}=100;\:n_{3,j}^{(2)}=50;\:n_{4,j}^{(2)}=100;\:n_{5,j}^{(2)}=300$ \\
			8 & $n_{1,j}^{(1)}=75;\:n_{2,j}^{(1)}=25;\:n_{3,j}^{(1)}=10;\:n_{4,j}^{(1)}=25;\:n_{5,j}^{(1)}=75$ & $n_{1,j}^{(2)}=150;\:n_{2,j}^{(2)}=50;\:n_{3,j}^{(2)}=20;\:n_{4,j}^{(2)}=50;\:n_{5,j}^{(2)}=150$ \\
			10 & $n_{1,j}^{(1)}=60;\:n_{2,j}^{(1)}=20;\:n_{3,j}^{(1)}=10;\:n_{4,j}^{(1)}=20;\:n_{5,j}^{(1)}=60$ & $n_{1,j}^{(2)}=120;\:n_{2,j}^{(2)}=40;\:n_{3,j}^{(2)}=20;\:n_{4,j}^{(2)}=40;\:n_{5,j}^{(2)}=120$ \\
			\bottomrule
		\end{tabular}
	\end{table}
\end{landscape}

\subsection{Results}\label{sec:results}

\begin{sidewaysfigure}
	\centerline{\includegraphics[width=18cm,height=6cm]{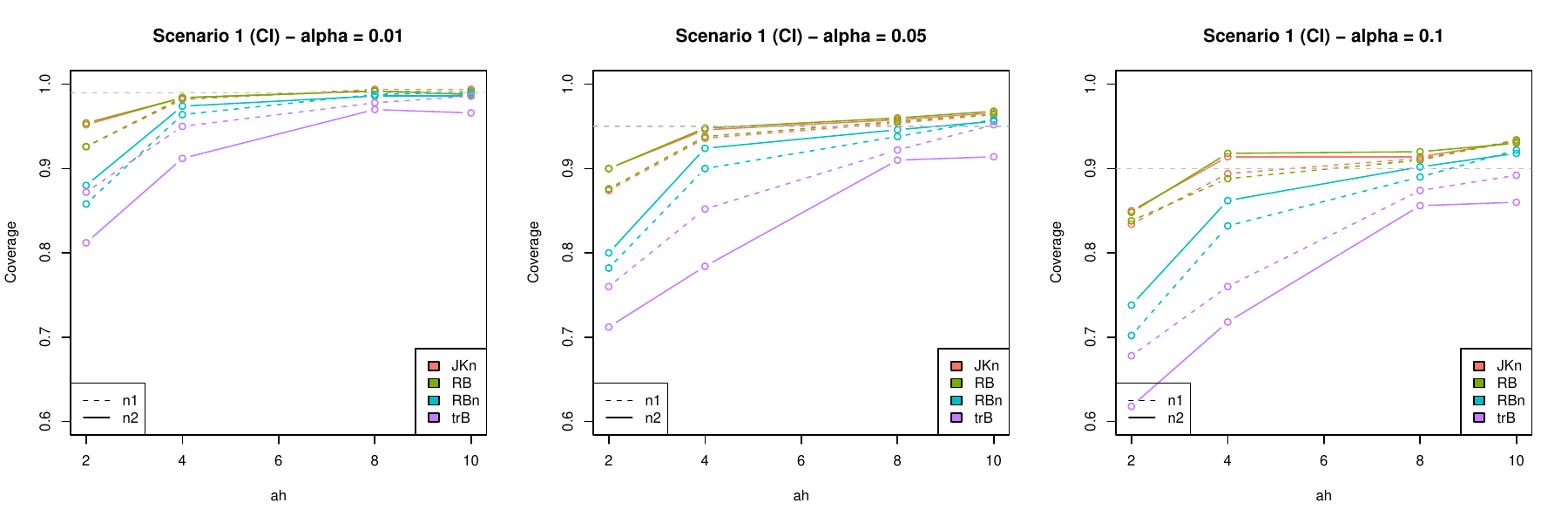}}
	\caption{Scenario 1 (confidence intervals). Coverage probabilities for nominal significance levels $\alpha\in\{0.01,0.05,0.1\}$ (from left to right). Results are shown for different numbers of clusters selected per stratum ($a_h\in\{2,4,8,10\}$) and for two total sample sizes ($n_1$ and $n_2$). The gray horizontal line represents the nominal coverage level ($1-\alpha$).\label{res:ci}}
	\centerline{\includegraphics[width=18cm,height=9cm]{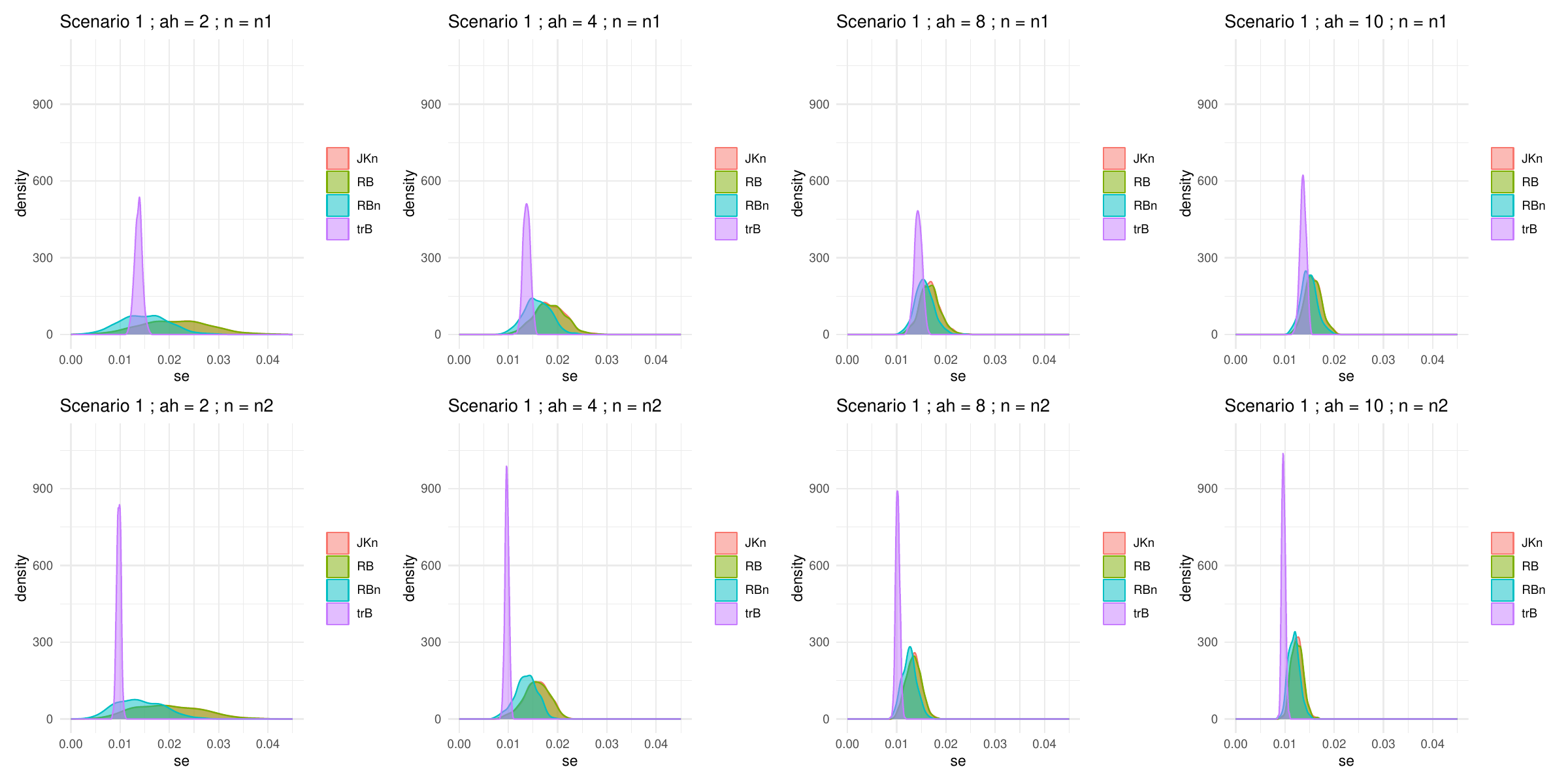}}
	\caption{Scenario 1 (confidence intervals). Estimated density functions of the standard errors for all methods. Columns correspond to different numbers of clusters selected per stratum ($a_h\in\{2,4,8,10\}$, from left to right), and rows correspond to the two total sample sizes considered ($n_1$ top, $n_2$ bottom).\label{res:sds-ci}}
\end{sidewaysfigure}

\begin{sidewaysfigure}
	\centerline{\includegraphics[width=18cm,height=6cm]{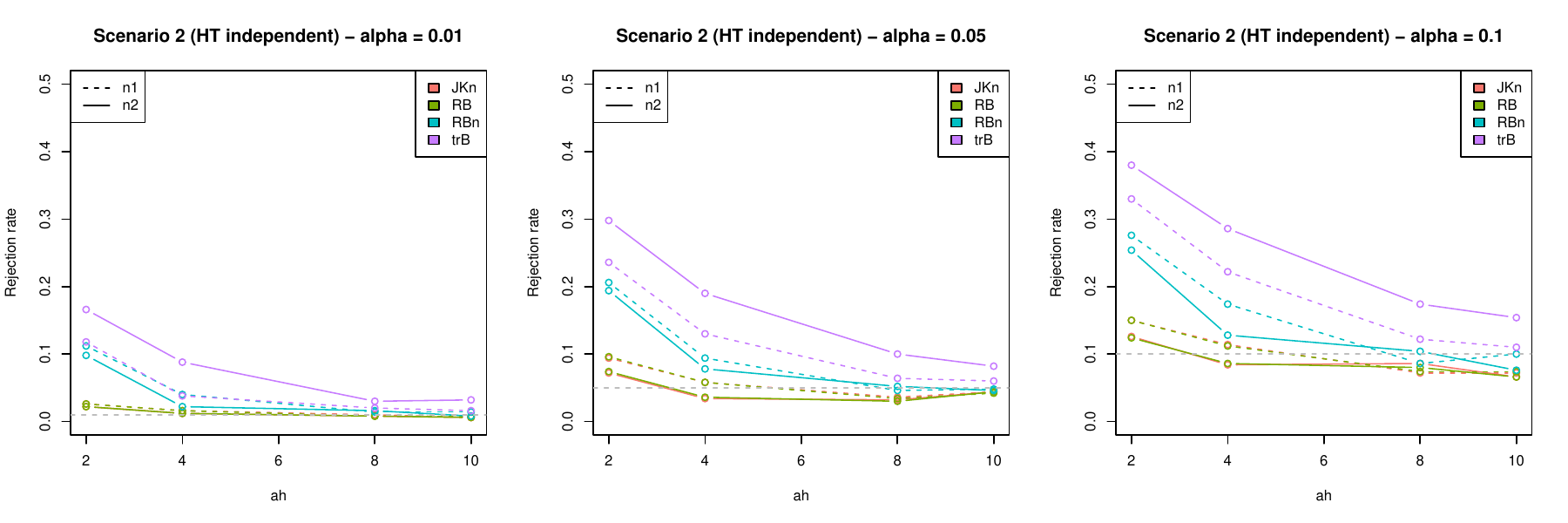}}
	\caption{Scenario 2 (hypothesis test for independent AUCs under $H_0$). Rejection rates for nominal significance levels $\alpha\in\{0.01,0.05,0.1\}$ (from left to right, indicated with the gray horizontal line). Results are shown for different numbers of clusters selected per stratum ($a_h\in\{2,4,8,10\}$) and for two total sample sizes ($n_1$ and $n_2$). \label{res:unpaired-eq}}
	%
	%\centerline{\includegraphics[width=18cm,height=9cm]{Figures/sds_unpaired_equal_ylim.pdf}}
	%\caption{Scenario 2 (hypothesis test for independent AUCs under $H_0$). Estimated density functions of the estimated standard errors for all methods. Columns correspond to different numbers of clusters selected per stratum ($a_h\in\{2,4,8,10\}$, from left to right), and rows correspond to the two total sample sizes considered ($n_1$ top, $n_2$ bottom).\label{res:sds-unpaired-eq}}
	%\end{sidewaysfigure}

	%\begin{sidewaysfigure}
	\centerline{\includegraphics[width=18cm,height=6cm]{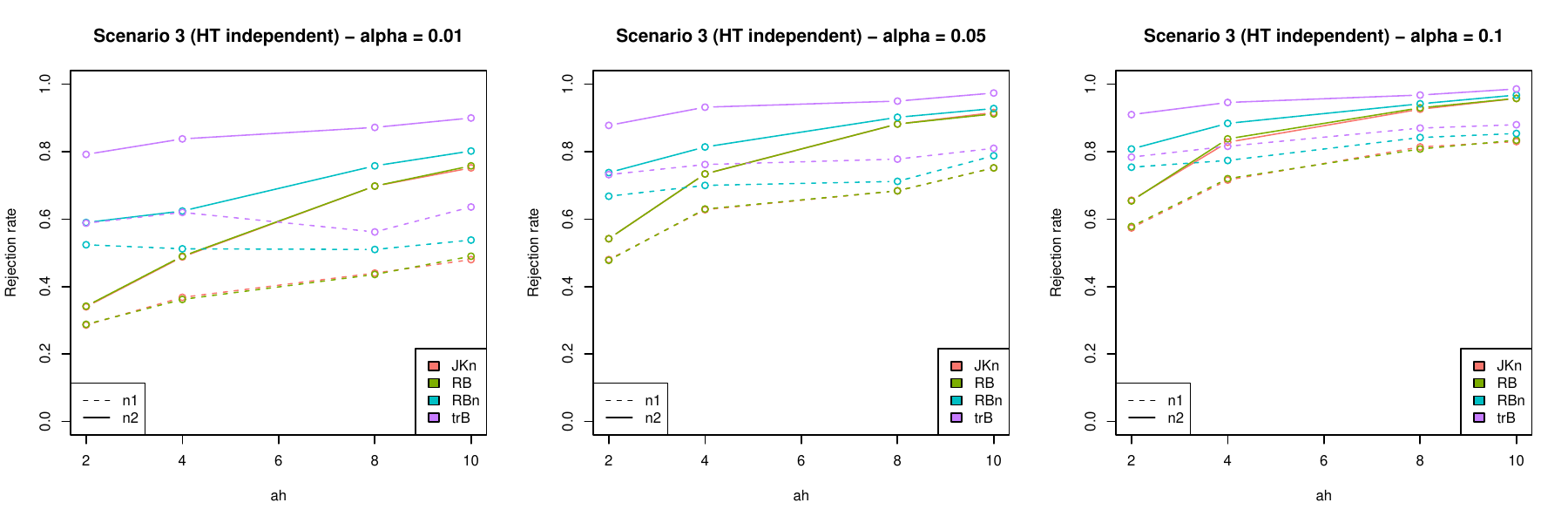}}
	\caption{Scenario 3 (hypothesis test for independent AUCs under $H_1$). Rejection rates for nominal significance levels $\alpha\in\{0.01,0.05,0.1\}$ (from left to right). Results are shown for different numbers of clusters selected per stratum ($a_h\in\{2,4,8,10\}$) and for two total sample sizes ($n_1$ and $n_2$). \label{res:unpaired-diff05}}
	%
	%\centerline{\includegraphics[width=18cm,height=9cm]{Figures/sds_unpaired_diff05_ylim.pdf}}
	%\caption{Scenario 3 (hypothesis test for independent AUCs under $H_1$). Estimated density functions of the estimated standard errors for all methods. %Columns correspond to different numbers of clusters selected per stratum ($a_h\in\{2,4,8,10\}$, from left to right), and rows correspond to the two total sample sizes considered ($n_1$ top, $n_2$ bottom).\label{res:sds-unpaired-diff05}}
\end{sidewaysfigure}

\begin{sidewaysfigure}
	\centerline{\includegraphics[width=18cm,height=6cm]{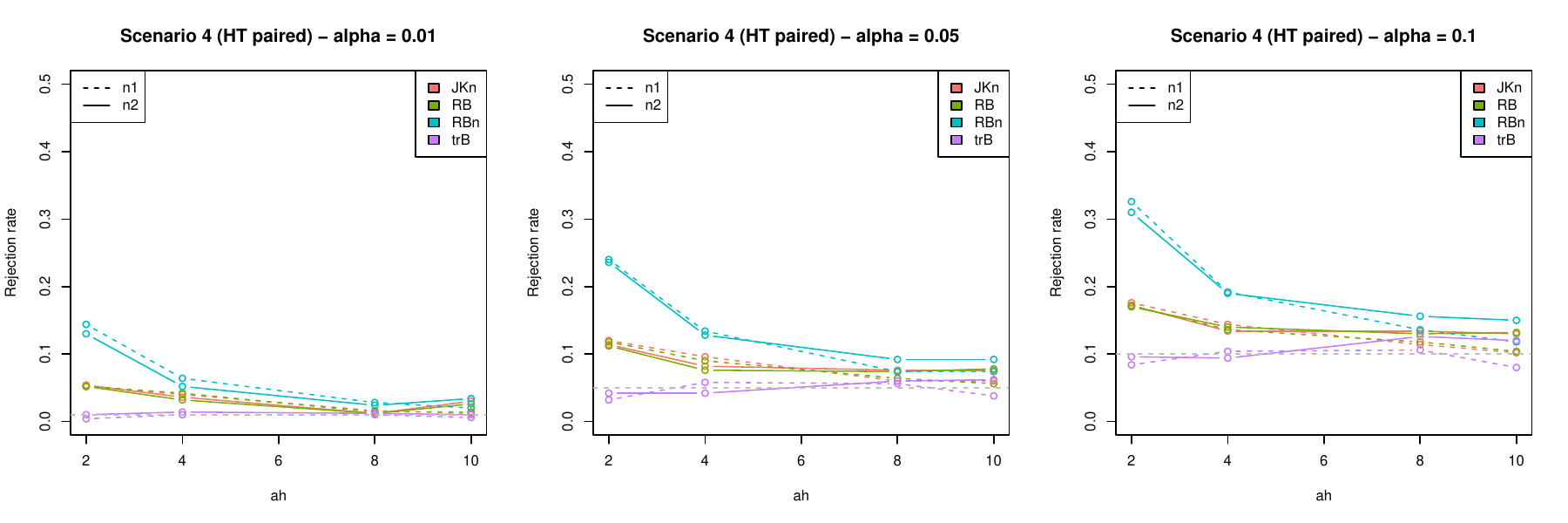}}
	\caption{Scenario 4 (hypothesis test for paired AUCs under $H_0$). Rejection rates for nominal significance levels $\alpha\in\{0.01,0.05,0.1\}$ (from left to right, indicated with the gray horizontal line). Results are shown for different numbers of clusters selected per stratum ($a_h\in\{2,4,8,10\}$) and for two total sample sizes ($n_1$ and $n_2$). \label{res:paired-eq}}
	%
	%\centerline{\includegraphics[width=18cm,height=9cm]{Figures/sds_paired_equal_ylim.pdf}}
	%\caption{Scenario 4 (hypothesis test for paired AUCs under $H_0$). Estimated density functions of the estimated standard errors for all methods. Columns correspond to different numbers of clusters selected per stratum ($a_h\in\{2,4,8,10\}$, from left to right), and rows correspond to the two total sample sizes considered ($n_1$ top, $n_2$ bottom).\label{res:sds-paired-eq}}
	%\end{sidewaysfigure}
	
	%\begin{sidewaysfigure}
	\centerline{\includegraphics[width=18cm,height=6cm]{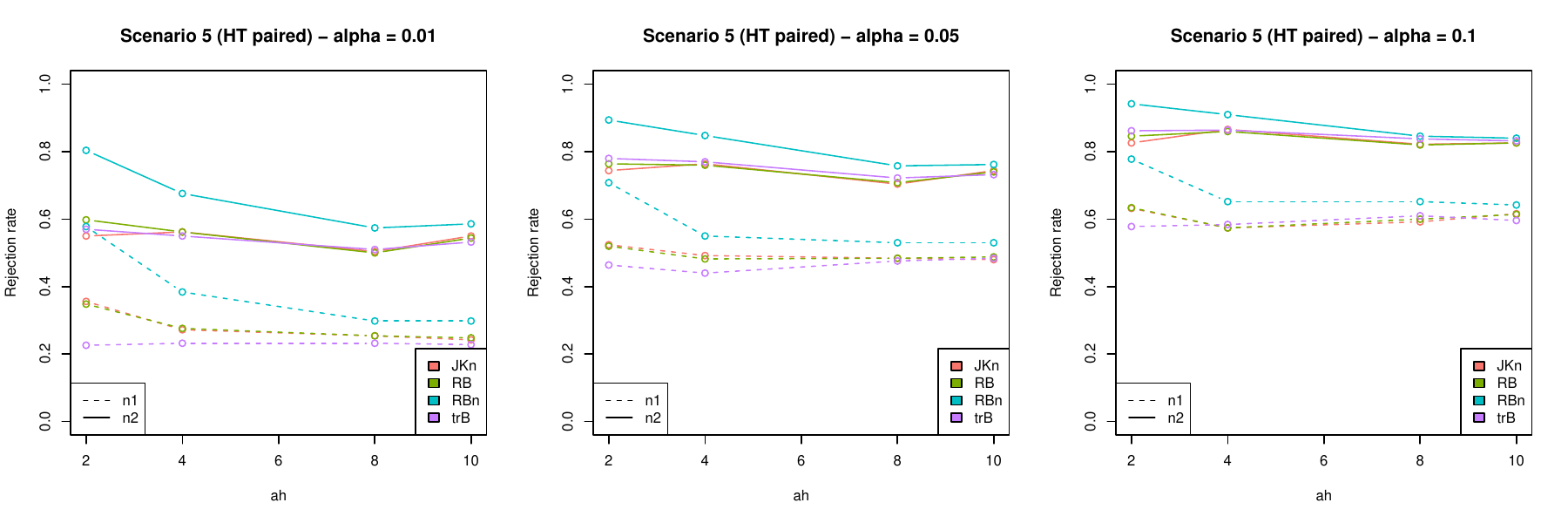}}
	\caption{Scenario 5 (hypothesis test for paired AUCs under $H_1$). Rejection rates for nominal significance levels $\alpha\in\{0.01,0.05,0.1\}$ (from left to right). Results are shown for different numbers of clusters selected per stratum ($a_h\in\{2,4,8,10\}$) and for two total sample sizes ($n_1$ and $n_2$).\label{res:paired-diff025}}
	%
	%\centerline{\includegraphics[width=18cm,height=9cm]{Figures/sds_paired_diff025_ylim.pdf}}
	%\caption{Scenario 5 (hypothesis test for paired AUCs under $H_1$). Estimated density functions of the estimated standard errors for all methods. Columns correspond to different numbers of clusters selected per stratum ($a_h\in\{2,4,8,10\}$, from left to right), and rows correspond to the two total sample sizes considered ($n_1$ top, $n_2$ bottom).\label{res:sds-paired-diff025}}
\end{sidewaysfigure}

\begin{sidewaysfigure}
	\centerline{\includegraphics[width=18cm,height=9cm]{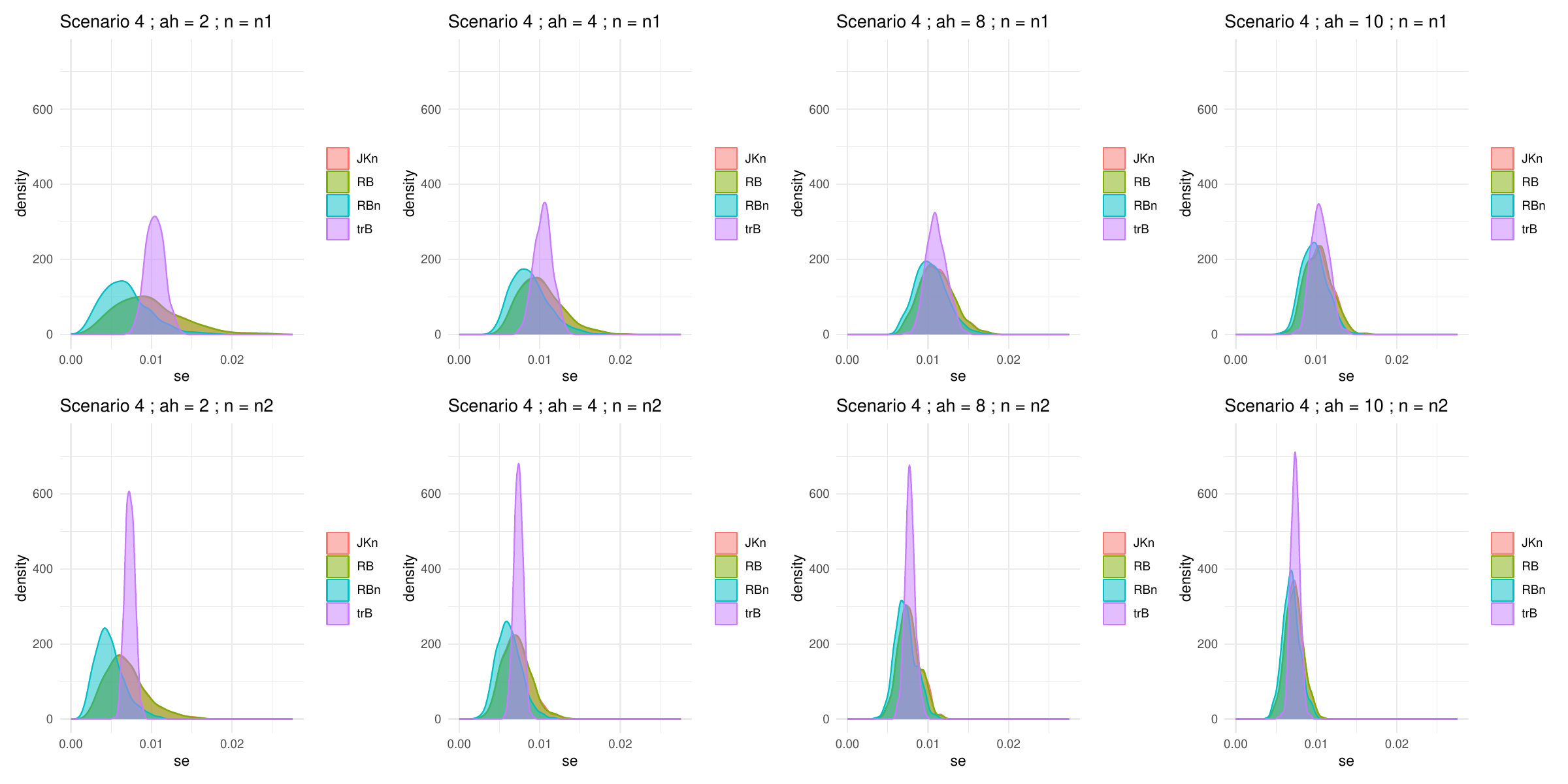}}
	\caption{Scenario 4 (hypothesis test for paired AUCs under $H_0$). Estimated density functions of the standard errors for all methods. Columns correspond to different numbers of clusters selected per stratum ($a_h\in\{2,4,8,10\}$, from left to right), and rows correspond to the two total sample sizes considered ($n_1$ top, $n_2$ bottom).\label{res:sds-paired-eq}}
\end{sidewaysfigure}

Figures~\ref{res:ci}–\ref{res:sds-paired-eq} present a summary of the results of the simulation study for all the methods and scenarios under consideration. For each scenario described in Table~\ref{tab:data_generation_summary}, the plots display the coverage probabilities of the confidence intervals (eq.~\eqref{eq:coverage}, Figure~\ref{res:ci}) and the rejection rates of the two-sided hypothesis tests (eq.~\eqref{eq:rejection_rate}, Figures~\ref{res:unpaired-eq}, \ref{res:unpaired-diff05}, \ref{res:paired-eq}, and \ref{res:paired-diff025}). Results are reported for significance levels $\alpha \in \{0.01, 0.05, 0.1\}$ (confidence levels $99\%$, $95\%$, and $90\%$, respectively), for different numbers of clusters selected per stratum ($a_h\in\{2,4,8,10\}$) and for two distinct total sample sizes ($n_1$ and $n_2$) as indicated in Table~\ref{tab:sampling}. Additionally, the estimated density functions of the square root of the variance estimators (i.e., the estimated standard errors) obtained with each method in Scenarios 1 and 4 are shown in Figures~\ref{res:sds-ci} and \ref{res:sds-paired-eq} , allowing a direct comparison of the distributional behaviour of the methods for variance estimation across sampling settings (to avoid redundancy and for space considerations, estimated standard errors corresponding to Scenarios 2, 3 and 5 are presented as Supplementary Information). To maintain consistency across the implemented methods, confidence intervals were constructed using the normal approximation as in eq.~\eqref{eq:CI_boot}. However, results obtained using the percentile bootstrap method in eq.~\eqref{eq:CI_boot_q} are provided as Supplementary Material.

Given the large number of results, we first summarize the main findings and then examine each scenario in greater detail. Overall, JKn and RB provide consistently satisfactory (and almost identical) results across all considered scenarios, both in terms of coverage probabilities and rejection rates with results close to nominal levels. Although RBn accounts for the sampling design, its performance is not appropriate, in particular, when only a small number of clusters per stratum are selected, with lower estimated standard errors leading to lower coverage probabilities and inflated rejection rates. The behaviour of trB varies depending on the scenario, but in general terms, it does not ensure reliable results across different scenarios except in the setting of the comparison of two paired AUCs. In the following lines, results are discussed for each scenario, jointly examining the coverage probabilities and rejection rates along with the distributional behaviour of the corresponding variance estimators.

In Scenario 1, the JKn and RB methods achieve the closest coverage probabilities to the nominal levels (Figure~\ref{res:ci}). Below them, RBn coverage lies between the higher-performing JKn and RB methods, and the lower-performing trB method. All methods get closer to the nominal levels and converge toward similar results as the number of clusters selected per stratum $a_h$ increases. Regarding the total sample size, JKn, RB, and RBn slightly improve their performance with larger samples, whereas trB performs worse as the sample size increases. These patterns are also reflected in the distributions of the estimated standard errors shown in Figure~\ref{res:sds-ci}. The trB method produces the smallest variance estimates (even smaller for larger sample sizes) leading to narrower confidence intervals and lower coverage, followed by RBn, and with RB and JKn yielding almost identical distributions with the largest estimated variances. Increasing the number of clusters per stratum reduces the differences between the methods, leading to more similar distributions of the estimated standard errors across approaches.

In Scenario 2, the performance of the methods in hypothesis testing for the comparison of independent AUCs is evaluated. Although the point estimates of the population AUCs differ slightly as shown in Table~\ref{tab:data_generation_summary} (0.7951 vs. 0.7941), this variation arises solely from the random generation of the populations using different seeds. All other parameters and covariate distributions were held constant (see Table~\ref{tab:data_generation_summary}), ensuring that the underlying discriminative ability of the models is effectively identical. Thus, in this scenario we assume that the null hypothesis is effectively true. Overall, the conclusions are similar to those seen for the confidence intervals, both in terms of rejection rate (Figure~\ref{res:unpaired-eq}) and in the distributional behaviour of the estimated standard errors (Figure S2 presented as Supplementary Material). Under the assumption that the null hypothesis is satisfied, the rejection rates of the two-sided hypothesis tests are expected to be close to the nominal significance levels. This is indeed observed for the JKn and RB methods, which produce rejection rates very close to the nominal $\alpha$ values. In contrast, rejection rates corresponding to RBn and trB tend to exceed the nominal levels, particularly for smaller numbers of clusters per stratum, and, in the case of trB, also for larger sample sizes, as noted previously in Scenario 1 for confidence intervals.

In Scenario 3, we analyze the performance of the methods for hypothesis testing of independent AUCs when the population AUCs are considerably different (0.7951 vs. 0.8474), so that the null hypothesis does not hold. In this setting, the rejection rate reflects the statistical power of the test, i.e., its ability to correctly reject the null hypothesis (Figure~\ref{res:unpaired-diff05}). As expected from previous observations, the highest power is achieved with trB and RBn, which tend to produce smaller standard error estimates leading to higher rejection rates. For all methods, an increase in the total sample size leads to a substantial gain in power. Additionally, selecting a larger number of clusters per stratum also improves the power of the test, particularly for the design-based RB and JKn methods. To analyze the effect of increasing the difference between the population AUCs, an additional scenario has been included as Supplementary Information, showing that the statistical power of the test increases accordingly (see Scenario 8 in Section S3 and Figure S6 provided as Supplementary Information).

In Scenario 4, regarding the performance of the methods on hypothesis testing for the comparison of two paired AUCs, we can assume that the null hypothesis holds and the minor differences in the finite population AUCs are negligible (0.7755 vs. 0.7743), as the covariates included in both models were generated to yield identical discriminative ability (see Table~\ref{tab:data_generation_summary}). In this case, the results differ slightly from those observed in previous scenarios. The trB method is the one that comes closest to the nominal significance levels, with values nearer than those obtained with RB and JKn (see Figure~\ref{res:paired-eq}). RBn is the method that differs the most from the nominal values with considerably higher rejection rates. The standard error estimates obtained with the replicate weight methods (JKn, RB and RBn) are generally smaller than those obtained by trB, particularly when a small number of clusters per stratum are selected (see Figure~\ref{res:sds-paired-eq}). As the number of clusters per stratum increases, all methods tend to behave very similarly, as also observed in previous scenarios.

In Scenario 5, the performance of the methods is evaluated for hypothesis testing for the comparison of two paired AUCs. In this scenario, the population AUCs differ (0.7991 vs. 0.8237), so the null hypothesis does not hold. Similar to Scenario 4, the smallest standard errors estimated with RBn, making this method the one with the highest power (Figure~\ref{res:paired-diff025}). For JKn, RB, and trB, the observed power is very similar across methods. Increasing the sample size substantially improves the power of the tests for all methods, whereas in this scenario no noticeable improvement in power is observed with an increasing number of clusters per stratum. As also observed in the comparison of independent AUCs, statistical power increases with larger differences in population AUCs (see Supplementary Information).

In summary, Scenarios 4 and 5, corresponding to the comparison of paired AUCs, are the only settings in which trB has shown good performance, outperforming the replicate-weight methods. To examine whether this behaviour is always observed when comparing paired AUCs or whether it is specific to the data-generating setting considered, we designed an additional scenario, which is described in Section~\ref{sec:additional_scenarios}.

\subsection{Extended scenarios for the comparison of two paired AUCs}\label{sec:additional_scenarios}

In this section, we present another setting for the comparison of two paried AUCs. The idea behind this additional scenario is the following. In previous scenarios, variables $X_3$ and $X_4$ were generated with the same variance–covariance structure with respect to the remaining variables (see eq.~\eqref{eq:var_covar}). As a consequence, including $X_3$ in one model and $X_4$ in the other in Scenarios 4 and 5 introduces essentially the same design-effect in both AUC estimates. When computing the difference between the paired AUCs, this common design-effect may be canceled. This could explain why trB, although being a non-design-based method, tends to produce rejection rates closer to the nominal level.

To analyze whether this phenomenon explains the behaviour observed in Scenarios 4 and 5, we modified the variance-covariance matrix so that the relationships of $X_3$ and $X_4$ with the remaining variables differ. In this way, the comparison of paired AUCs no longer removes the full impact of the sampling design. Specifically, the variance–covariance matrix previously defined in eq.~\eqref{eq:var_covar} was modified so that the covariances between $X_3$ and the remaining variables were set to 0.5, except for $X_4$, with which the covariance was set to 0. In addition, $X_4$ was defined to be uncorrelated with all the other variables. To construct two distinct scenarios (one with equal population AUCs and one with different population AUCs) the mean vectors previously defined in Table~\ref{tab:data_generation_summary} were modified as indicated in Table~\ref{tab:diffvar}.
%\begin{center}
\begin{table}[h!]%
	\centering
	\footnotesize
	\caption{Summary of two more scenarios considered in the simulation study for the comparison of paired AUCs. For each scenario, the mean vector $\pmb\mu_{Y=1}^T$ of the covariates for units with $Y=1$, the covariates considered in the model, and the corresponding population AUC are reported.}\label{tab:diffvar}
	\begin{tabular}{cccccc}
		\toprule
		Scenario & Contrast & Population & $\pmb\mu_{Y=1}^T$ & Covariates & $AUC^{\text{pop}}$\\
		\midrule
		6 & HT & $(1)$  &  $(0.7,0.7,1,0.5,0.7,0.7,0.7,0.7,0.7,0.7)$  &  $X_1,X_2,X_3$  & 0.7735\\
		&paired &   &    &  $X_1,X_2,X_4$  & 0.7732\\
		\midrule
		7 & HT & $(1)$  &  $(0.7,0.7,1,0.2,0.7,0.7,0.7,0.7,0.7,0.7)$  &  $X_1,X_2,X_3$  & 0.7735\\
		&paired &   &   &  $X_1,X_2,X_4$  & 0.7493 \\
		\bottomrule
	\end{tabular}
\end{table}
%\end{center}

\begin{sidewaysfigure}
		\centering
	\includegraphics[width=18cm,height=6cm]{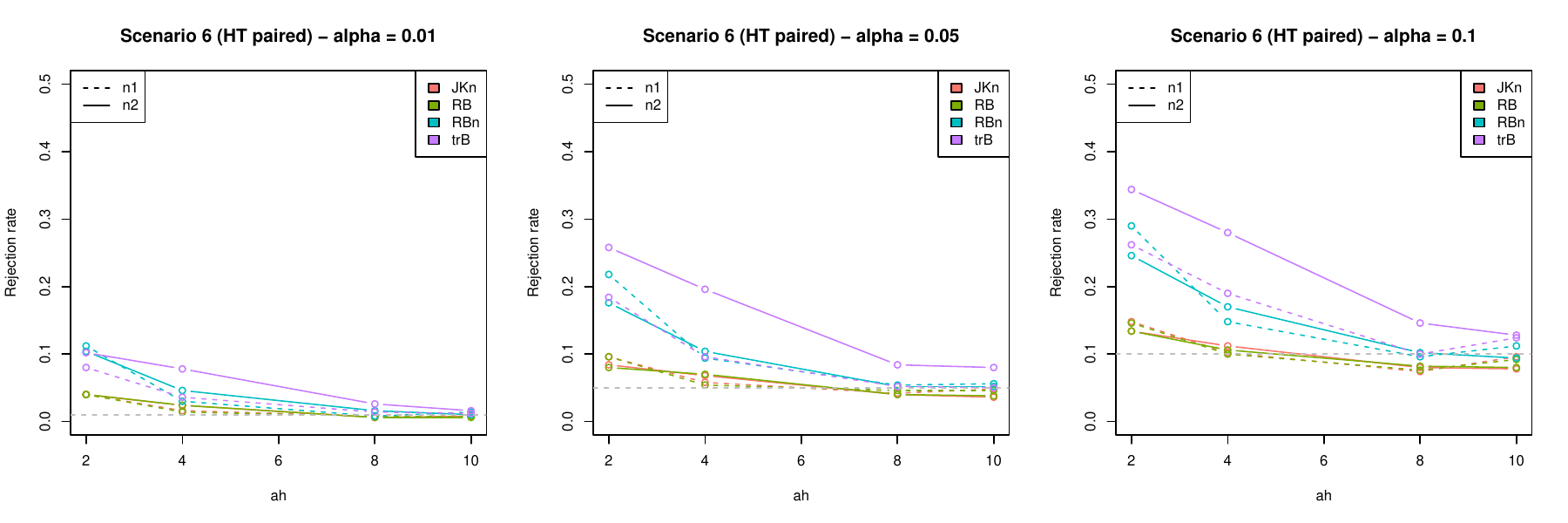}
	\caption{Scenario 6 (hypothesis test for paired AUCs under $H_0$). Estimated density functions of the standard errors for all methods. Columns correspond to different numbers of clusters selected per stratum ($a_h\in\{2,4,8,10\}$, from left to right), and rows correspond to the two total sample sizes considered ($n_1$ top, $n_2$ bottom).}
	\label{res:paired-eq-diffvar}
	%
	%\centerline{\includegraphics[width=18cm,height=9cm]{Figures/sds_paired_equal_diffvar_xlim.pdf}}
	%\caption{Scenario 6 (hypothesis test for paired AUCs under $H_0$). Estimated density functions of the  estimated standard errors for all methods. Columns correspond to different numbers of clusters selected per stratum ($a_h\in\{2,4,8,10\}$, from left to right), and rows correspond to the two total sample sizes considered ($n_1$ top, $n_2$ bottom).\label{res:sds-paired-eq-diffvar}}
	%\end{sidewaysfigure}
	
	%\begin{sidewaysfigure}
	\includegraphics[width=18cm,height=6cm]{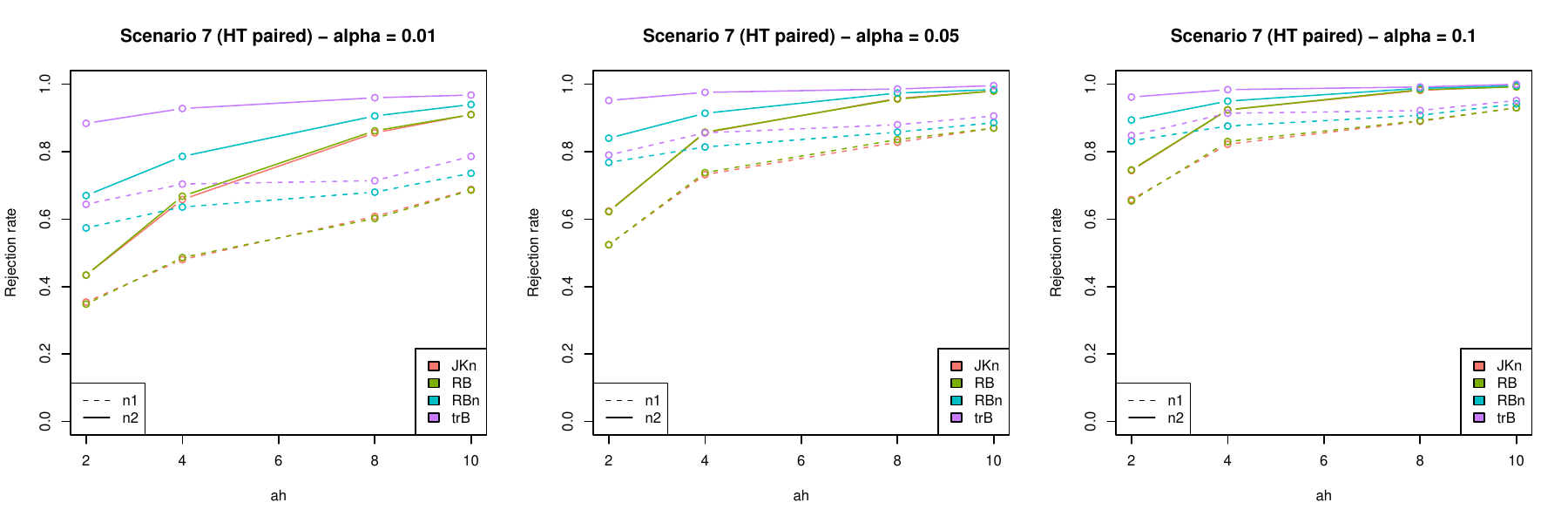}
	\caption{Scenario 7 (hypothesis test for paired AUCs under $H_1$). Rejection rates for nominal significance levels $\alpha\in\{0.01,0.05,0.1\}$ (from left to right). Results are shown for different numbers of clusters selected per stratum ($a_h\in\{2,4,8,10\}$) and for two total sample sizes ($n_1$ and $n_2$).}
	\label{res:paired-uneq-diffvar}
	%
	%\centerline{\includegraphics[width=18cm,height=9cm]{Figures/sds_paired_diffvar_025_xlim.pdf}}
	%\caption{Scenario 7 (hypothesis test for paired AUCs under $H_1$). Estimated density functions of the  estimated standard errors for all methods. Columns correspond to different numbers of clusters selected per stratum ($a_h\in\{2,4,8,10\}$, from left to right), and rows correspond to the two total sample sizes considered ($n_1$ top, $n_2$ bottom).\label{res:sds-paired-uneq-diffvar}}
\end{sidewaysfigure}

\begin{sidewaysfigure}
	\centering
	\includegraphics[width=18cm,height=9cm]{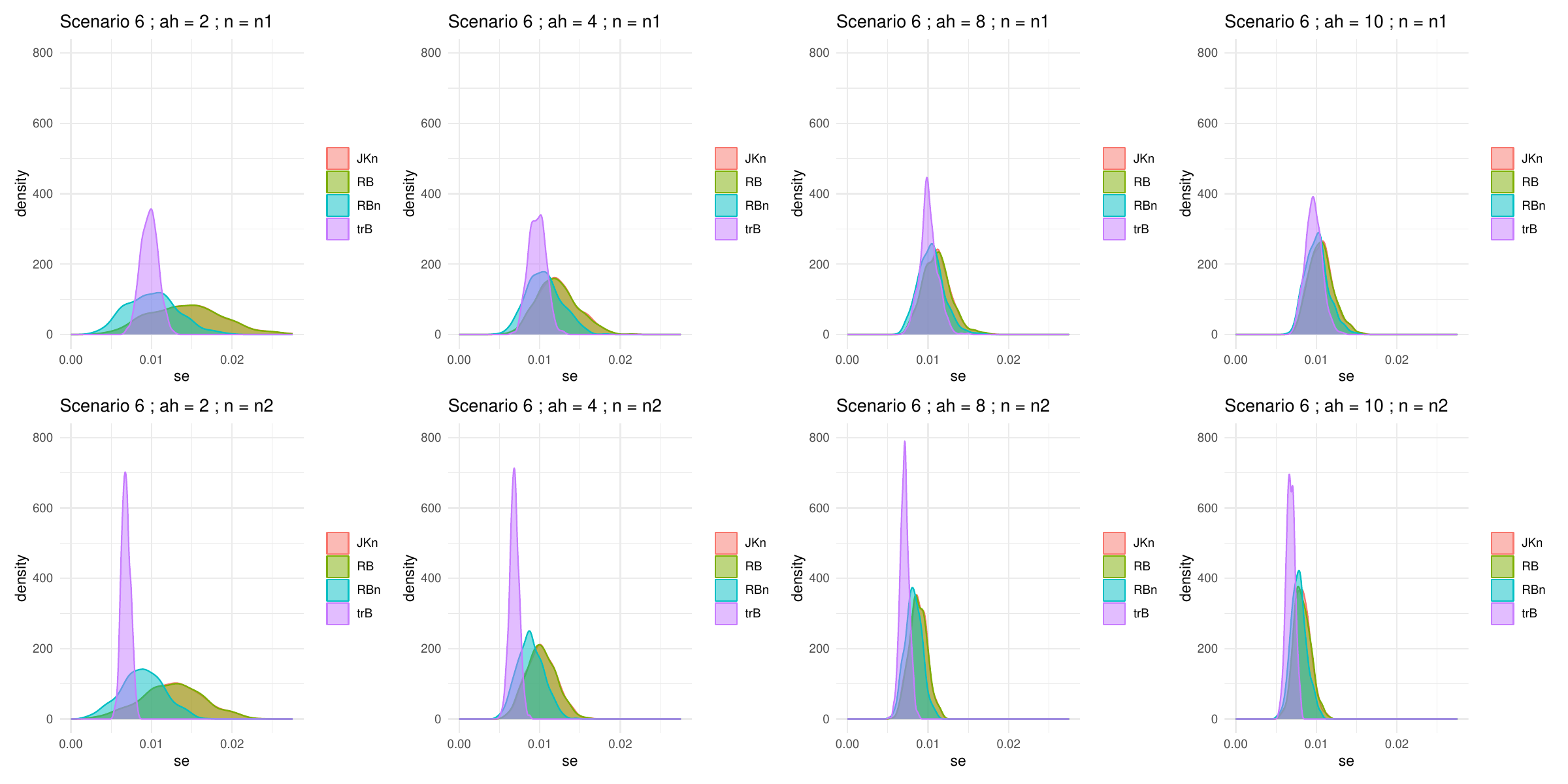}
	\caption{Scenario 6 (hypothesis test for paired AUCs under $H_0$). Estimated density functions of the  estimated standard errors for all methods. Columns correspond to different numbers of clusters selected per stratum ($a_h\in\{2,4,8,10\}$, from left to right), and rows correspond to the two total sample sizes considered ($n_1$ top, $n_2$ bottom).}
	\label{res:sds-paired-eq-diffvar}
\end{sidewaysfigure}

Figures~\ref{res:paired-eq-diffvar} and \ref{res:paired-uneq-diffvar} depict the results in terms of rejection rates for Scenarios 6 and 7, respectively. The estimated standard errors for Scenario 6 are shown in Figure~\ref{res:sds-paired-eq-diffvar} (similar distributions were obtained for Scenario 7 and are displayed in Figure S5 provided as Supplementary Information). The results show that, in these scenarios, trB produces smaller variance estimates than JKn and RB, which translates into larger rejection rates that deviate more from the nominal significance levels. These results are similar to those obtained in Scenarios 2 and 3 (see Figures~\ref{res:unpaired-eq} and \ref{res:unpaired-diff05}). Thus, we conclude that the performance of trB is scenario-dependent and may not be reliable, even in the context of paired AUC comparisons.

\section{Application}\label{sec:application}

In this section, the methodology proposed and evaluated in the previous sections is applied to a real survey data. It should be emphasized that the purpose of this application is purely illustrative. The analysis was restricted to complete cases for simplicity and comparability across models. No attempt was made to address missing data through more advanced techniques, since the primary goal is not to draw epidemiological conclusions but the illustration of the proposed methodology using real survey data. Similarly, we have not carried out any formal model selection procedure, although all included covariates are statistically significant ($\alpha=0.05$). Consequently, the conclusions drawn from this application are not intended to have clinical implications; rather, the focus is exclusively on the statistical behaviour of the proposed methods and on AUC inference under complex sampling designs.

Data from the National Health and Nutrition Examination Survey (NHANES) was considered. NHANES is a national survey that measures the health and nutrition of adults and children in the United States which includes health exams, laboratory tests and dietary interviews for participants of all ages. This survey is conducted on a continuous basis and data are released in two-year cycles. It employs a complex and multistage sampling design involving stratification and clustering. 

In particular, data from the 2021–2023 NHANES cycle (August 2021 to August 2023) was considered for the analysis (dataset $(1)$, hereinafter). This dataset includes $11\,933$ sampled individuals, distributed across 15 strata with two primary sampling units per stratum. Diabetes status (dichotomous) was used as the response variable, and age, gender, educational level, poverty income ratio (PIR), and body mass index (BMI) were included as covariates. The analysis was restricted to complete cases resulting in a final sample of $5\,002$ individuals. For the comparison of independent AUCs, data from the 2011–2012 NHANES cycle was additionally considered (dataset $(2)$, hereinafter). This dataset contains $9\,756$ individuals, distributed across 14 strata with two or three primary sampling units per stratum. After restricting to complete cases for the selected variables, the final sample size is $4\,699$ individuals. Prevalence is 0.119 in dataset (1) and 0.093 in dataset (2).

Table~\ref{tab:nhanes_models} summarizes the covariates included in the five models considered in this application (M1–M5). Specifically, one analysis was conducted for confidence interval estimation (model M1, using dataset (1)), two analyses were performed for the comparison of paired AUCs (comparing models M2–M3 and models M4–M5, both based on dataset (1)), and one analysis was carried out for the comparison of independent AUCs (fitting model M1 separately in datasets (1) and (2)). The corresponding results are presented in Table~\ref{tab:nhanes_results}. 95\% confidence intervals are defined for all the methods considered (JKn, RB, RBn and trB) and hypothesis tests are interpreted with significance level $\alpha=0.05$. The complex sampling design was considered throughout the whole analysis.

The results show that the estimated confidence intervals are very similar across all methods, with the only noticeable difference being that RBn yields a slightly narrower interval than the remaining approaches. Regarding the first paired AUC comparison (models M2 and M3, with point estimates 0.798 and 0.802, respectively), RBn rejects the null hypothesis of equal AUCs at the 5\% significance level, whereas the other methods do not. It is also worth noting that the largest estimated standard error is obtained using trB. These findings are consistent with those observed in Scenarios 4 and 5 of the simulation study. For the second paired AUC comparison (models M4 and M5, with point estimates 0.709 and 0.760, respectively), all methods reject the null hypothesis of equal AUCs. In this case, the largest estimated standard errors are obtained using RB and JKn, in line with the behaviour observed in Scenarios 6 and 7. Finally, for the comparison of independent AUCs (model M1 fitted in dataset (1), with estimated AUC 0.804, and in dataset (2), with estimated AUC 0.828), RB and JKn again produce the largest estimated standard errors and therefore the largest p-values. RBn yields smaller variance estimates, while trB provides intermediate results. These findings are comparable to those observed in Scenarios 2 and 3. Overall, the results from the NHANES application are consistent with the patterns previously identified in the simulation study. 

%\begin{center}
\begin{table*}[!h]%
	\centering
	\caption{Covariates included in each of the five models considered in the analysis conducted using the NHANES data.\label{tab:nhanes_models}}
	\begin{tabular*}{\textwidth}{@{\extracolsep\fill}lccccc@{}}
		\toprule
		Covariates & M1 & M2 & M3 & M4 & M5\\
		\midrule
		Age & $\checkmark$ & $\checkmark$ & $\checkmark$ & & $\checkmark$\\
		Gender & $\checkmark$ & $\checkmark$ & $\checkmark$ & $\checkmark$ & $\checkmark$ \\
		Education level & $\checkmark$ & & $\checkmark$ & $\checkmark$ & $\checkmark$\\
		PIR & $\checkmark$ & $\checkmark$ & & $\checkmark$ & $\checkmark$ \\
		BMI & $\checkmark$ & $\checkmark$ & $\checkmark$ & $\checkmark$ &\\
		\bottomrule
	\end{tabular*}
	%\begin{tablenotes}%%[341pt]
	%\item[$^{\rm a}$] Example for a first table footnote.
	%\item[$^{\rm b}$] Example for a second table footnote.
	%\item {\it Source}: Example for table source text.
	%\end{tablenotes}
\end{table*}
%\end{center}
%\begin{center}
\begin{landscape}
\begin{table*}[!h]%
	\centering
	\footnotesize
	\caption{Summary of all analyses conducted using the NHANES data and their corresponding numerical results.\label{tab:nhanes_results}}
	\begin{tabular*}{\textwidth}{@{\extracolsep\fill}lc|cccccc|cccccc|cccccc@{}}
		\cmidrule{1-20}
		& CI       & \multicolumn{6}{c|}{HT paired}                                                                                & \multicolumn{6}{c|}{HT paired}                                                                                & \multicolumn{6}{c}{HT indep}                                                                                 \\
		%\cmidrule{2-2}\cmidrule{3-8}\cmidrule{9-14}\cmidrule{15-20}
		\cmidrule{1-20}
		Dataset           & (1)      &\multicolumn{3}{c}{(1)}                                          & \multicolumn{3}{c|}{(1)}                   & \multicolumn{3}{c}{(1)}                                          & \multicolumn{3}{c|}{(1)}                   & \multicolumn{3}{c}{(1)}                                          & \multicolumn{3}{c}{(2)}                   \\
		Model(s)          & M1       & \multicolumn{3}{c}{M2}                                           & \multicolumn{3}{c|}{M3}                    & \multicolumn{3}{c}{M4}                                           & \multicolumn{3}{c|}{M5}                    & \multicolumn{3}{c}{M1}                                           & \multicolumn{3}{c}{M1}                    \\
		$\widehat{AUC}_w$ &     0.804     & \multicolumn{3}{c}{0.798}                                             & \multicolumn{3}{c|}{0.802}                      & \multicolumn{3}{c}{0.709}                                             & \multicolumn{3}{c|}{0.760}                      & \multicolumn{3}{c}{0.804}                                             & \multicolumn{3}{c}{0.828}                      \\
		%\cmidrule{2-2}\cmidrule{3-8}\cmidrule{9-14}\cmidrule{15-20}
		\cmidrule{1-20}
		& $95\%$ CI & \multicolumn{2}{c}{se} & \multicolumn{2}{c}{$z_m$} & \multicolumn{2}{c|}{p-value} & \multicolumn{2}{c}{se} & \multicolumn{2}{c}{$z_m$} & \multicolumn{2}{c|}{p-value} & \multicolumn{2}{c}{se} & \multicolumn{2}{c}{$z_m$} & \multicolumn{2}{c}{p-value} \\
		%\cmidrule{2-2}\cmidrule{3-8}\cmidrule{9-14}\cmidrule{15-20}
		\cmidrule{1-20}
		JKn               &    $(0.786,\:0.822)$      &     \multicolumn{2}{c}{0.002}                     &      \multicolumn{2}{c}{-1.455}                   &    \multicolumn{2}{c|}{0.146}         &     \multicolumn{2}{c}{0.017}        &      \multicolumn{2}{c}{-2.953}         &     \multicolumn{2}{c|}{0.003}        &           \multicolumn{2}{c}{0.017}               &     \multicolumn{2}{c}{-1.422}                    &    \multicolumn{2}{c}{0.155}         \\
		RB                &     $(0.786,\:0.823)$      &     \multicolumn{2}{c}{0.002}                     &      \multicolumn{2}{c}{-1.483}                   &     \multicolumn{2}{c|}{0.138}        &      \multicolumn{2}{c}{0.017}       &      \multicolumn{2}{c}{-2.971}         &      \multicolumn{2}{c|}{0.003}       &         \multicolumn{2}{c}{0.017}                 &      \multicolumn{2}{c}{-1.418}                   &    \multicolumn{2}{c}{0.156}         \\
		RBn               &    $(0.791,\:0.817)$       &    \multicolumn{2}{c}{0.002}                      &    \multicolumn{2}{c}{ -2.130}                    &      \multicolumn{2}{c|}{0.033}       &     \multicolumn{2}{c}{0.012}        &     \multicolumn{2}{c}{-4.198}          &    \multicolumn{2}{c|}{$<0.001$}         &     \multicolumn{2}{c}{0.013}                     &    \multicolumn{2}{c}{-1.892}                     &    \multicolumn{2}{c}{0.058}         \\
		trB               &     $(0.785,\:0.824)$      &     \multicolumn{2}{c}{0.004}                     &      \multicolumn{2}{c}{-0.993}                  &     \multicolumn{2}{c|}{0.321}        &     \multicolumn{2}{c}{0.014}        &       \multicolumn{2}{c}{-3.599}        &       \multicolumn{2}{c|}{$<0.001$}      &      \multicolumn{2}{c}{0.014}                    &      \multicolumn{2}{c}{-1.644}                   &     \multicolumn{2}{c}{0.100}        \\
		\cmidrule{1-20}
	\end{tabular*}
\end{table*}
\end{landscape}
%\end{center}

\section{Conclusions}\label{sec:conclusions}

This article introduces a design-based framework for inference on the AUC in complex survey data. The proposed methodology can be used to construct confidence intervals and to perform hypothesis tests when comparing independent or paired AUCs. Specifically, it relies on replicate weights methods to estimate the variance of the weighted AUC estimator. An extensive simulation study was conducted, highlighting the importance of properly accounting for the sampling design, particularly when the number of PSUs per stratum is limited.

Among the methods considered, JKn and RB showed the best overall performance, performing well across all scenarios considered. The variance estimates obtained with these two methods were almost identical, with the only noticeable difference between both methods being computational cost. In the simulation study and the NHANES application analyzed in this work, JKn was considerably faster than the bootstrap approach. Nevertheless, it should be noted that the computational cost of JKn increases with the number of PSUs, so in settings with a large number of PSUs the bootstrap may become more computationally efficient.

The results also show that both the number of PSUs and the overall sample size have a consistent impact on inference quality. In general, increasing the number of PSUs per stratum and the total sample size leads to improved coverage and higher statistical power across all methods. Conversely, when the number of PSUs is limited, differences between methods become more pronounced and the impact of variance underestimation when the sampling design is not properly taken into account becomes more evident.

The practical relevance of the proposed methodology is illustrated through its application to the NHANES survey. The empirical results are consistent with the patterns observed in the simulation study. Although the purpose of this application is purely illustrative rather than clinical, it demonstrates that the proposed framework performs as expected under real survey conditions and can be readily implemented in widely used population-based health studies.

The proposed framework can also be extended to other AUC estimators, such as the one based on pairwise sampling weights considering joint inclusion probabilities~\citep{Yao2015}. Although the impact of missing data was not specifically investigated in this work, evaluating how different missing data strategies interact with design-based AUC inference is an interesting topic to evaluate in future research~\citep{Martins2025}. In addition, other concepts related to the AUC, such as the covariate adjustments~\citep{Janes2008,Inacio2022}, are also intended as topics for future research.

All the code implemented in the simulation study and the NHANES application are publicly available on GitHub\footnote{\url{https://github.com/aiparragirre/design-based-AUC-inference/tree/main}}, ensuring transparency and reproducibility. Furthermore, the methods proposed in this work have been incorporated into the \texttt{svyROC} R package, facilitating their practical implementation in real survey data. Together, these resources provide a robust and accessible framework for design-based AUC inference in complex survey settings.

%Pairwise weights-ekin ere izan daiteke. Guk marginal erabili dugu. Bien arteko konparaketa, aurrerago.

%\begin{sidewaysfigure}
%\centerline{\includegraphics[width=542pt,height=9pc,draft]{empty}}
%\caption{Sideways figure caption. Sideways figure caption. Sideways figure caption. Sideways figure caption. Sideways figure caption. Sideways figure caption.\label{fig3}}
%\end{sidewaysfigure}

%\backmatter
%\bmsection*{Author contributions}

%This is an author contribution text. This is an author contribution text. This is an author contribution text. This is an author contribution text. This is an author contribution text.

%\bmsection*{Acknowledgments}
%This is acknowledgment text. Provide text here. This is acknowledgment text. Provide text here. This is acknowledgment text. Provide text here. This is acknowledgment text. Provide text here. This is acknowledgment text. Provide text here. This is acknowledgment text. Provide text here. This is acknowledgment text. Provide text here. This is acknowledgment text. Provide text here. This is acknowledgment text. Provide text here.

\subsubsection*{Funding}

This work was supported by the University of the Basque Country through POSTUPV24/58 and EHU-N25/19, by a grant from the Department of Science, University and Innovation from the Basque Government to the MATHMODE Group (IT1866-26), by the Spanish Ministry of Science and Innovation through BCAM Severo Ochoa accreditation [CEX2021–001142-S / MICIN / AEI /10.13039/501100011033], through the project PID2024-156800OB-I00 funded by Agencia Estatal de Investigación and acronym ``STARHS'', and through RED2024-153680-T (BIOSTATNET) funded by MICIU/AEI /10.13039/501100011033; by the Basque Government through the BERC 2022-2025 program, and by the Instituto de Salud Carlos III (ISCIII) through the project RD24/0005/0020  (Red de Investigación en Cronicidad, Atención Primaria y Prevención y Promoción de la Salud).\\

\subsubsection*{Conflict of interest}

The authors declare no potential conflict of interests.

\bibliographystyle{chicago}
\bibliography{mybib}

\end{document}